\documentclass[reprint,amsmath,amssymb,amsfonts,aps,pre,floatfix]{revtex4-1}

\usepackage{amsmath}
\usepackage{bm}
\usepackage{cancel}
\usepackage{graphicx}
\usepackage{subfigure}
\usepackage{wrapfig}
\usepackage{ifpdf}
\usepackage{rotating}
\usepackage{alltt}
\usepackage{float}
\usepackage{fullpage}
\usepackage{color}
\usepackage[usenames,dvipsnames]{xcolor}

\usepackage{times}
\DeclareMathSizes{10}{10}{8}{8}
\usepackage{geometry}
\newcommand{\abovesectionspace}{\vspace{-0.2in}}
\newcommand{\belowsectionspace}{\vspace{-0.1in}}
\newcommand{\abovesubsectionspace}{\vspace{-0.2in}}
\newcommand{\belowsubsectionspace}{\vspace{-0.2in}}
\usepackage{fancyhdr}
\usepackage{afterpage}
\usepackage{parskip}
\newcommand{\nl}[1]{\right.\\#1\left.{}}
\newcommand{\circlenum}[1]{\raisebox{.5pt}{\textcircled{\raisebox{-.9pt} {\textit{#1}}}}}

\newcommand{\figspace}{\vspace{-0.1in}}

\DeclareMathOperator*{\define}{\equiv}

\newcommand{\eq}[1]{ Eq.\ (\ref{#1})}

\DeclareMathOperator*{\Stresstype}{\boldsymbol{\Pi}}
\DeclareMathOperator*{\Stresstypekinetic}{\boldsymbol{\kappa}}

\DeclareMathOperator*{\PressureVA}{\boldsymbol{\Stresstype\limits^{\scriptscriptstyle{V\!A}}}}
\DeclareMathOperator*{\PressureVAkinetic}{\boldsymbol{\Stresstypekinetic\limits^{\scriptscriptstyle{V\!A}}}}
\DeclareMathOperator*{\Stresstypecauchy}{\boldsymbol{\sigma}}

\DeclareMathOperator*{\StressVAcauchy}{\Stresstypecauchy\limits^{\scriptscriptstyle{VA}}}
\DeclareMathOperator*{\rhouutype}{\{\rho \textbf{u} \textbf{u}\}}
\DeclareMathOperator*{\rhouuVA}{\boldsymbol{\rhouutype\limits^{\scriptscriptstyle{V \! A}}}}

\newcommand{\CV}{$\mathcal{LCV}$}

\begin{document}
\title{$ \; $ \\
\vspace{0.2in}
\large{Control-volume representation of molecular dynamics}}


\author{E.~R.~Smith, D.~M.~Heyes, D.~Dini and T.~A.~Zaki}
\affiliation{Department of Mechanical Engineering, Imperial College London, Exhibition Road, London SW7 2AZ, United Kingdom} 
\email[]{edward.smith05@imperial.ac.uk; \; \\ 
d.heyes@imperial.ac.uk; \; \\ 
d.dini@imperial.ac.uk; \; \\ 
t.zaki@imperial.ac.uk; \;}


\date{Received 13 October 2011; revised manuscript received 2 March 2012; published 22 May 2012}

\begin{abstract}

A Molecular Dynamics (MD) parallel to the Control Volume (CV) formulation of fluid mechanics is developed by integrating the formulas of Irving and Kirkwood, J. Chem. Phys. 18, 817 (1950)  over a finite cubic volume of molecular dimensions.
The Lagrangian molecular system is expressed in terms of an Eulerian CV, which yields an equivalent to Reynolds' Transport Theorem for the discrete system.
This approach casts the dynamics of the molecular system into a form that can be readily compared to the continuum equations.  
The MD equations of motion are reinterpreted in terms of a Lagrangian-to-Control-Volume (\CV) conversion function $\vartheta_{i}$, for each molecule $i$.
The \CV \ function and its spatial derivatives are used to express fluxes and relevant forces across the control surfaces.  
The relationship between the local pressures computed using the Volume Average (VA, Lutsko, J. Appl. Phys 64, 1152 (1988) ) techniques and the Method of Planes (MOP , Todd et al, Phys.
Rev. E 52, 1627 (1995) 
) emerges naturally from the treatment. 
Numerical experiments using the MD CV method are reported for equilibrium and 
non-equilibrium (start-up Couette flow) model liquids, which demonstrate the advantages of the formulation. The CV formulation of the MD is shown to be exactly conservative, and is therefore ideally suited to obtain macroscopic properties from a discrete system. 

\vspace{0.2in}
\noindent
\textbf{DOI:} 10.1103/PhysRevE.85.056705 \;\;\;\;\;\;\;\;\;\;\;\;\;\;\;\;\;\;\;\;\;\;\;\;\; PACS number(s): 05.20.−y, 47.11.Mn, 31.15.xv

\end{abstract}



\maketitle

\lhead{\; }
\chead{\small{PHYSICAL REVIEW E 85, 056705 (2012)}}
\rhead{\; }
\lfoot{1539-3755/2012/85(5)/056705(19)}
\cfoot{056705-\thepage}
\renewcommand{\headrulewidth}{0.0pt}
\renewcommand{\footrulewidth}{0.0pt}
\thispagestyle{fancy}

\abovesectionspace
\section{Introduction}
\belowsectionspace

The macroscopic and microscopic descriptions of mechanics have traditionally been studied independently.  
The former invokes a continuum assumption, and aims to reproduce the large-scale behaviour of solids and fluids, without the need to resolve the micro-scale details.  
On the other hand, molecular simulation predicts the evolution of individual, but interacting, molecules, which has application in nano and micro-scale systems. 
Bridging these scales requires a mesoscopic description, which represents the evolution of the average of many microscopic trajectories through phase space.  
It is advantageous to cast the fluid dynamics equations in a consistent form for both the molecular, mesoscale and continuum approaches.  
The current works seeks to achieve this objective by introducing a Control Volume (CV) formulation for the molecular system.  

The Control Volume approach is widely adopted in continuum fluid mechanics, where Reynolds Transport Theorem \citep{Reynolds} relates Newton's laws of motion for macroscopic fluid parcels to fluxes through a CV. 
In this form, fluid mechanics has had great success in simulating both fundamental 
\citep{Zaki_Durbin_05,Zaki_Durbin_06} and practical \citep{Hirsch,Rosenfeld_et_al,Zaki_Durbin_10} flows. 
However, when the continuum assumption fails, or when macroscopic constitutive equations are lacking, a molecular-scale description is required.  
Examples include nano-flows, moving contact lines, solid-liquid boundaries, non-equilibrium fluids, and evaluation of transport properties such as viscosity and heat conductivity \citep{Evans_Morris}.

Molecular Dynamics (MD) involves solving Newton's equations of motion for an assembly of interacting discrete molecules. 
Averaging is required in order to compute properties of interest, e.g.\ temperature, density, pressure and stress, which can vary on a local scale especially out of equilibrium \citep{Evans_Morris}. 
A rigorous link between mesoscopic and continuum properties was established in the seminal work of \citet{Irving_Kirkwood}, who related the mesoscopic Liouville equation to the differential form of continuum fluid mechanics. 
However, the resulting equations at a point were expressed in terms of the Dirac $\delta$ function --- a form which is difficult to manipulate and cannot be applied directly in a molecular simulation. 
Furthermore, a Taylor series expansion of the Dirac $\delta$ functions was required to express the pressure tensor. 
The final expression for pressure tensor is neither easy to interpret nor to compute \citep{Zhou}. 
As a result, there have been numerous attempts to develop an expression for the pressure tensor for use in MD simulation \citep{Parker, Noll, Tsai, Todd_et_al_95, Han_Lee, Hardy, Lutsko, Cormier_et_al, Zhou, Murdoch, Murdoch_2010, Schofield_Henderson, Admal_Tadmor}. 
Some of these expressions have been shown to be equivalent in the appropriate limit.  For example, \citet{Heyes_et_al}) demonstrated equivalence between Method of Planes (MOP \citet{Todd_et_al_95}) and Volume Average (VA \citet{Lutsko}) at a surface.

In order to avoid use of the Dirac $\delta$ function, the current work adopts a Control Volume representation of the MD system, written in terms of fluxes and surface stresses. 
This approach is in part motivated by the success of the control volume formulation in continuum fluid mechanics.
At a molecular scale, control volume analyses of NEMD simulations can facilitate evaluation of local fluid properties.  
Furthermore, the CV method also lends itself to coupling schemes between the continuum and molecular descriptions \citep{ OConnell_Thompson, Hadjiconstantinou_thesis, Li_et_al, Hadjiconstantinou_99, Flekkoy_et_al, Wagner_et_al, Delgado-Buscalioni_Coveney_03, Curtin_Miller, Nie_et_al, Werder_et_al, Ren, Borg_et_al}.

The equations of continuum fluid mechanics are presented in Section \ref{sec:Continuum_Equations}, followed by a review of the \citet{Irving_Kirkwood} procedure for linking  continuum and mesoscopic properties in Section \ref{sec:IK}.
In section \ref{sec:CV}, a Lagrangian to Control Volume (\CV) conversion function is used to express the mesoscopic equations for mass and momentum fluxes.
Section \ref{sec:pressure} focuses on the stress tensor, and relates the current formulation to established definitions within the literature \citep{Lutsko, Cormier_et_al,Todd_et_al_95}. 
In Section \ref{sec:implement}, the CV equations are derived for a single microscopic system, and subsequently integrated in time in order to obtain a form which can be applied in MD simulations.  
The conservation properties of the CV formulation are demonstrated in NEMD simulations of Couette flow in Section \ref{sec:MD_results_discussion}. 

\afterpage{\cfoot{056705-\thepage}}
\afterpage{\lhead{\small{E. R. SMITH, D. M. HEYES, D. DINI, AND T. A. ZAKI} \\}}
\afterpage{\chead{\;}}
\afterpage{\rhead{\small{PHYSICAL REVIEW E 85, 056705 (2012)}\\}}
\afterpage{\lfoot{\;}}
\renewcommand{\headrulewidth}{0.0pt}
\renewcommand{\footrulewidth}{0.0pt}
\thispagestyle{fancy}

\abovesectionspace
\section{Background}
\belowsectionspace

This section summarizes the theoretical background. First, the macroscopic continuum equations are introduced, followed by the mesoscopic equations which describe the evolution of an ensemble average of systems of discrete molecules. The link between the two descriptions is subsequently discussed.

\abovesubsectionspace
\subsection{Macroscopic Continuum Equations}
\belowsectionspace
\label{sec:Continuum_Equations}

The continuum conservation of mass and momentum balance can be derived in an Eulerian frame by considering the fluxes through a Control Volume (CV).  The mass continuity equation can be expressed as,
\begin{align}
	\frac{\partial }{\partial t}  \int_V  \rho dV = - \oint_S \rho \boldsymbol{u} \cdot d\textbf{S},
\label{BofmEqn2}
\end{align}
where $\rho$ is the mass density and $\boldsymbol{u}$ is the fluid velocity. The rate of change of momentum is determined by the balance of forces on the CV,
\begin{align}
	\!\! \frac{\partial }{\partial t}  \int_V  \rho \boldsymbol{u} dV \! = - \oint_S \rho \boldsymbol{u} \boldsymbol{u} \cdot d\textbf{S}+ \textbf{F}_{\textnormal{surface}}  + \textbf{F}_{\textnormal{body}}.
\label{BofMEqn2}
\end{align}
The forces are split into ones which act on the bounding surfaces, $\textbf{F}_{\textnormal{surface}}$, and body forces, $\textbf{F}_{\textnormal{body}}$. 
Surface forces are expressed in terms the pressure tensor, $\boldsymbol{\Pi}$, on the CV surfaces,
\begin{align}
	\textbf{F}_{\textnormal{surface}} = - \oint_S \boldsymbol{\Pi} \cdot d\textbf{S}.
\label{Ftostress}
\end{align}
The rate of change of energy in a CV is expressed in terms of fluxes, the pressure tensor and a heat flux vector $\textbf{q}$, 
\begin{align}
	\!\! \frac{\partial }{\partial t}  \int_V  \rho \mathcal{E} dV \! = - \oint_S \left[ \rho \mathcal{E} \boldsymbol{u} +   \boldsymbol{\Pi} \cdot \boldsymbol{u}  +  \textbf{q} \right] \cdot d\textbf{S},
\label{BofenergyEqn}
\end{align}
here the energy change due to body forces is not included. The divergence theorem relates surface fluxes to the divergence within the volume, for a variable $A$,
 \begin{align}
\oint_S \boldsymbol{A}  \cdot d\textbf{S} = \int_V \boldsymbol{\nabla} \cdot \boldsymbol{A}  dV 
 \label{divergence}
 \end{align}
In addition, the differential form of the flow equations can be recovered in the limit of an infinitesimal control volume \citep{Borisenko_Tarapov},
 \begin{align}
\boldsymbol{\nabla} \cdot \boldsymbol{A}  = \lim_{V \rightarrow 0} \frac{1}{V}\oint_S \boldsymbol{A}   \cdot d\textbf{S}.  
\label{definition_of_grad}
 \end{align}

\abovesubsectionspace
\subsection{Relationship Between the Continuum and the Mesoscopic Descriptions}
\belowsubsectionspace
\label{sec:IK}

A mesoscopic description is a temporal and spatial average of the molecular trajectories, expressed in terms of a probability function, $\textit{f}$. 
\citet{Irving_Kirkwood} established the link between the mesoscopic and continuum descriptions using the Dirac $\delta$ function to define the macroscopic density at a point $\textbf{r}$ in space,
\begin{align}
\rho(\textbf{r},t) \define \displaystyle\sum_{i=1}^{N} \bigg\langle m_i  \delta(\textbf{r}_i-\textbf{r}) ; \textit{f} \bigg\rangle.
\label{massdensity}
\end{align}
The angled brackets $ \langle \alpha ; f \rangle $ denote the inner product of $\alpha$ with $\textit{f}$, which gives the expectation of $\alpha$ for an ensemble of systems. The mass and position of a molecule $i$ are denoted $m_i$ and $\textbf{r}_i$, respectively, and $N$ is the number of molecules in a single system. 
The momentum density at a point in space is similarly defined by,
\begin{align}
\rho(\textbf{r},t) \boldsymbol{u}(\textbf{r},t) \define \displaystyle\sum_{i=1}^{N} \bigg\langle \textbf{p}_i  \delta(\textbf{r}_i-\textbf{r}) ; \textit{f} \bigg\rangle,
\label{Momdensity}
\end{align}
where the molecular momentum, $\textbf{p}_i = m_i \dot{\textbf{r}}_i$. 
Note that $\textbf{p}_i$ is the momentum in the laboratory frame, and not the peculiar value $\overline{\textbf{p}}_i$ which excludes the macroscopic streaming term at the location of molecule $i$, $\boldsymbol{u}(\textbf{r}_i)$, \citep{Evans_Morris},
\begin{align}
\overline{\textbf{p}}_i \define m_i \left(\frac{\textbf{p}_i}{m_i} - \boldsymbol{u}(\textbf{r}_i) \right).
\label{peculiar}
\end{align} 
The present treatment uses $\textbf{p}_i$ in the lab frame.  A discussion of translating CV and its relationship to the peculiar momentum is given in Appendix \ref{sec:divergence_ReynoldsTT}. 

Finally, the energy density at a point in space is defined by
\begin{align}
\rho(\textbf{r},t) \mathcal{E}(\textbf{r},t) \define \displaystyle\sum_{i=1}^{N} \bigg\langle  e_i  \delta(\textbf{r}_i-\textbf{r}) ; \textit{f} \bigg\rangle,
\label{Energydensity}
\end{align}
where the energy of the $i^{th}$ molecule is defined as the sum of the kinetic energy and the inter-molecular interaction potential $\phi_{ij}$,
\begin{align}
 e_i \define \frac{p_i^2}{2m_i} +   \frac{1}{2}\displaystyle\sum_{j \ne i}^N \phi_{ij} 
\end{align}
It is implicit in this definition that the potential energy of an interatomic interaction, $\phi_{ij}$, is divided equally between the two interacting molecules, $i$ and $j$.

As phase space is bounded, the evolution of a property, $\alpha$, in time is governed by the equation,
 \begin{align}
\frac{\partial }{\partial t} \bigg\langle \alpha ; \textit{f} \bigg\rangle  = \displaystyle\sum_{i = 1 }^{N} \bigg\langle \textbf{F}_i \cdot \frac{\partial \alpha}{\partial \textbf{p}_i} +\frac{\textbf{p}_i}{m_i} \cdot \frac{\partial \alpha}{\partial \textbf{r}_i} ;\textit{f} \bigg\rangle,
\label{EqIK18}
\end{align}
where $\textbf{F}_i$ is the force on molecule $i$, and $\alpha=\alpha(\textbf{r}_i(t), \textbf{p}_i(t))$ is an implicit function of time.  Using Eq. (\ref{EqIK18}), \citet{Irving_Kirkwood} derived the time evolution of the mass (from Eq.\ \ref{massdensity}), momentum density (from Eq.\ \ref{Momdensity}) and energy density (from Eq.\ \ref{Energydensity}) for a mesoscopic system. 
A comparison of the resulting equations to the continuum counterpart provided a term-by-term equivalence.
Both the mesoscopic and continuum equations were valid at a point;  the former expressed in terms of Dirac $\delta$ and the latter in differential form.
In the current work, the mass and momentum densities are recast within the CV framework which avoids use of the Dirac $\delta$ functions directly, and attendant problems with their practical implementation.

\abovesectionspace
\section{The Control Volume Formulation}
\belowsectionspace
\label{sec:CV}

\begin{figure*}
  \subfigure{\label{fig:cube}\includegraphics[width=0.25\textwidth]{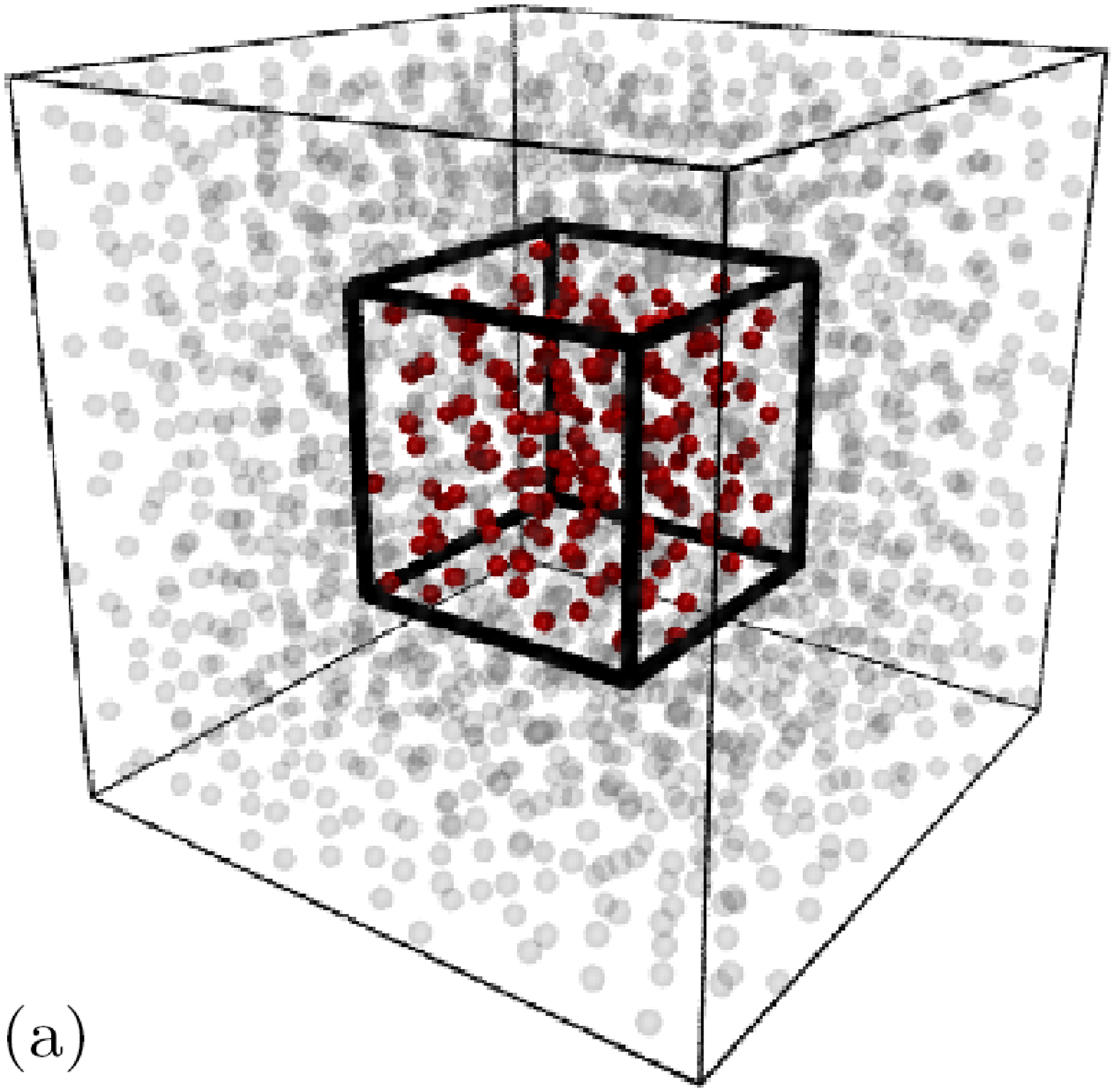}} \;\;\;\;\;\;\;\;\;
  \subfigure{\label{fig:axis}\includegraphics[width=0.25\textwidth]{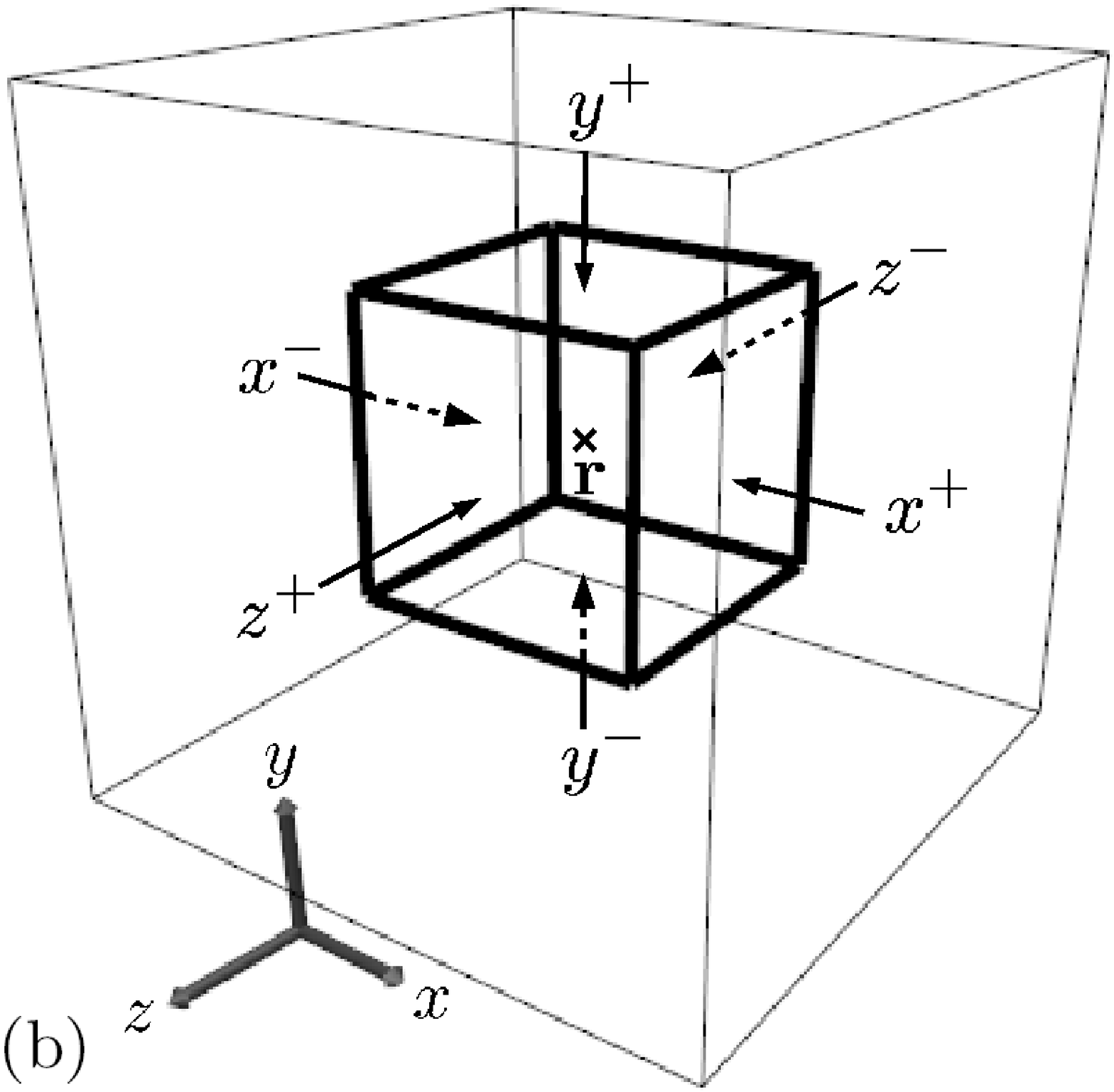}} \;\;\;\;\;\;\;\;\;
  \subfigure{\label{fig:planes}\includegraphics[width=0.25\textwidth]{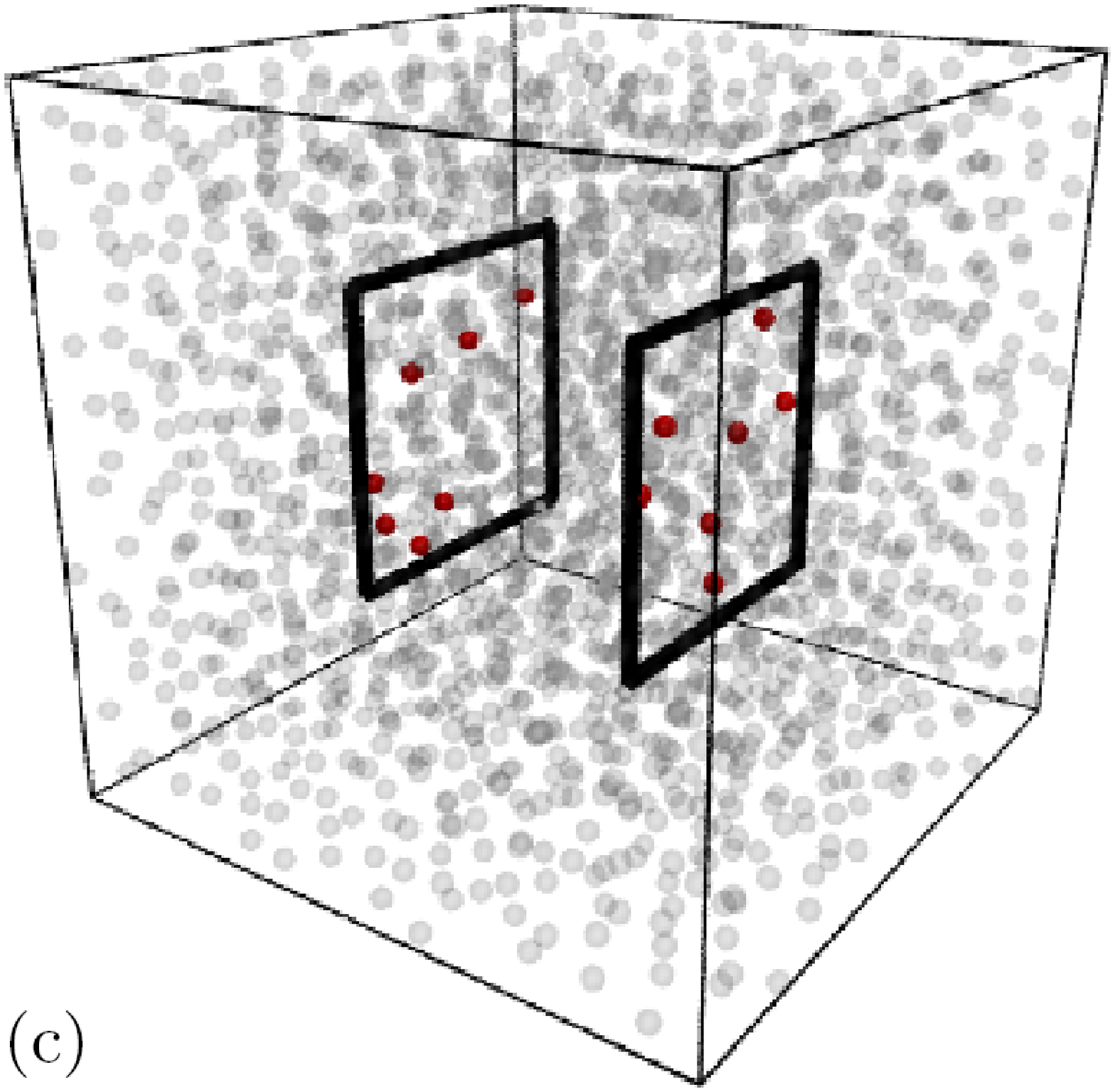}}
\caption{(Color online) The CV function and its derivative applied to a system of molecules. The figures were 
generated using the VMD visualization package, \citep{vmd}. From left to right, (a) Schematic of $\vartheta_i$ which selects only the molecules within a cube, (b) Location of cube center $\textbf{r}$ and labels for cube surfaces, (c) Schematic of $\partial \vartheta_i/\partial x$ which selects only molecules crossing the $x^+$ and $x^-$ surface planes.}
\figspace
\end{figure*}

In order to cast the governing equations for a discrete system in CV form, a `selection function' $\vartheta_i$ is introduced, which isolates those molecules within the region of interest. This function is obtained by integrating the Dirac $\delta$ function, $\delta(\textbf{r}_i - \textbf{r})$, over a cuboid in space, centered at $\textbf{r}$ and of side length $\Delta \textbf{r}$ as illustrated in figure \ref{fig:cube}
\footnote{The cuboid is chosen as the most commonly used shape in continuum mechanic simulations on structured grids, although the process could be applied to any arbitrary shape}.  
Using $\delta(\textbf{r}_i-\textbf{r}) = \delta(x_i-x)\delta(y_i-y)\delta(z_i-z)$, the resulting triple integral is,
\begin{align}
	\vartheta_i \define \int\limits_{x^-}^{x^+} \int\limits_{y^-}^{y^+} \int\limits_{z^-}^{z^+} \delta(x_i-x)\delta(y_i-y)\delta(z_i-z) dxdydz \; \; \; \;  \nonumber \\
	\!\!\! =\bigg[\bigg[\bigg[ H(x_i-x)H(y_i-y)H(z_i-z) \bigg]_{x^-}^{x^+} \bigg]_{y^-}^{y^+} \bigg]_{z^-}^{z^+}	\nonumber \\
	\!\!\! = \left[ H(x^+-x_i) - H(x^- - x_i) \right]  \; \,  \;\;\;\;\;\;\;\;\;\;\;\;\;\;\;\;\;		\nonumber \\
	\times \left[ H(y^+ - y_i) \; \!  - \, H(y^- - y_i) \right] \; \, \;\;\;\;\;\;\;\;\;\;\;\;\;\;\;\;\;	\nonumber \\
	\times \left[ H(z^+ - z_i) \, - \, H(z^- - z_i) \right], \;\;\;\;\;\;\;\;\;\;\;\;\;\;\;\;\;
\label{CV_function}
\end{align}
where $H$ is the Heaviside function, and the limits of integration are defined as, $\textbf{r}^- \define \textbf{r}-\frac{\Delta \textbf{\textbf{r}}}{2}$ and $\textbf{r}^+ \define \textbf{r}+\frac{\Delta \textbf{\textbf{r}}}{2}$, for each direction (see Fig. \ref{fig:axis}).
Note that $\vartheta_i$ can be interpreted as a Lagrangian-to-Control-Volume conversion function (\CV) \~ for molecule $i$.  
It is unity when molecule $i$ is inside the cuboid, and equal to zero otherwise, as illustrated in Fig.\ \ref{fig:cube}. 
Using  L'H\^{o}pital's rule and defining, $\Delta V \define \Delta x\Delta y\Delta z$, the \CV  \ function for molecule $i$ reduces to the Dirac $\delta$ function in the limit of zero volume,
\begin{align}
	\delta(\textbf{r}-\textbf{r}_i) = \lim_{\Delta V \rightarrow 0} \frac{\vartheta_i}{\Delta V}.  \nonumber 
\end{align}
The spatial derivative in the $x$ direction of the \CV \ function for molecule $i$ is,
\begin{align}
	\frac{\partial \vartheta_i}{\partial x} = -\frac{\partial \vartheta_i}{\partial x_i} =  \left[ \delta(x^+ - x_i)-\delta(x^- - x_i) \right] S_{xi},
\label{difftheta}
\end{align}
where $S_{xi}$ is
\begin{align}
	S_{xi} \define 	\left[ H(y^+ - y_i) \; \!  - \, H(y^- - y_i) \right] \; \,		\nonumber \\
				\left[ H(z^+ - z_i) \, - \, H(z^- - z_i) \right].
\label{Six}
\end{align}
 Eq.\ (\ref{difftheta}) isolates molecules on a 2D rectangular patch in the $yz$ plane. The derivative $\partial \vartheta_i/\partial x$ is only non-zero when molecule $i$ is crossing the surfaces marked in Fig.\ \ref{fig:planes}, normal to the $x$ direction.
The contribution of the $i^{th}$ molecule to the net rate of mass flux through the control surface is expressed in the form,  $\textbf{p}_i \cdot d\textbf{S}_i$.
Defining for the right $x$ surface,
\begin{align}
	dS_{xi}^+ \define \delta(x^+ - x_i) S_{xi},
\label{flux}
\end{align}
and similarly for the left surface, $dS_{xi}^-$, the total flux Eq.\ (\ref{difftheta}) in any direction $\textbf{r}$ is then,
 \begin{align}
	\frac{\partial \vartheta_i}{\partial \textbf{r}} =  d\textbf{S}_i^+ - d\textbf{S}_i^- \define d\textbf{S}_i.
\label{dSi}
\end{align}
The \CV \ function is key to the derivation of a molecular-level equivalent of the continuum CV equations, and it will be used extensively in the following sections. The approach in sections \ref{sec:ddtmass}, \ref{sec:ddtmomentum} and \ref{sec:ddtenergy} shares some similarities with the work of \citet{Serrano_Espanol} which considers the time evolution of Voronoi characteristic functions. However the \CV \ function has precisely defined extents which allows the development of conservation equations for a microscopic system. In the following treatment, the CV is fixed in space (i.e., $\textbf{r}$ is not a function of time). 
The extension of this treatment to an advecting CV is made in Appendix \ref{sec:divergence_ReynoldsTT}. 

\abovesubsectionspace
\subsection{Mass Conservation for a Molecular CV}
\belowsubsectionspace
\label{sec:ddtmass}
In this section, a mesoscopic expression for the mass in a cuboidal CV is derived. The time evolution of mass within a CV is shown to be equal to the net mass flux of molecules across its surfaces.

The mass inside an arbitrary CV at the molecular scale can be expressed in terms of the \CV \ as follows,
\begin{align}
	\int_V \rho(\textbf{r},t) dV = 
		\int_V \displaystyle\sum_{i=1}^{N}  \bigg\langle m_i  \delta(\textbf{r}_i-\textbf{r}) ; \textit{f} \bigg\rangle dV \;\;		\nonumber \\
		=  \displaystyle\sum_{i=1}^{N} \int\limits_{x^-}^{x^+} \!\! \int\limits_{y^-}^{y^+} \!\! \int\limits_{z^-}^{z^+} \!\! \bigg\langle m_i \delta(\textbf{r}_i-\textbf{r}) ; \textit{f} \bigg\rangle dxdydz
\nonumber \\
		\!=\! \displaystyle\sum_{i=1}^{N} \bigg\langle m_i  \vartheta_i ; \textit{f} \bigg\rangle.
\label{ID:mCFD2MD}
\end{align}
Taking the time derivative of Eq.\ (\ref{ID:mCFD2MD}) and using Eq.\ (\ref{EqIK18}),
\begin{align}
	\frac{\partial}{\partial t} \int_V  \rho(\textbf{r},t)  dV = \frac{\partial}{\partial t} \displaystyle\sum_{i=1}^{N}  \bigg\langle m_i \vartheta_i ; \textit{f} \bigg\rangle 	\nonumber \\ 
	= \displaystyle\sum_{i = 1 }^{N}  \bigg\langle \frac{\textbf{p}_{i}}{m_i} \cdot \frac{\partial}{\partial \textbf{r}_{i}} m_i \vartheta_i + \textbf{F}_{i} \cdot \frac{\partial}{\partial \textbf{p}_{i}} m_i \vartheta_i ; \textit{f}  \bigg\rangle.
\label{EqIK20}
\end{align}
The term $\partial m_i \vartheta_i /\partial \textbf{p}_{i} = 0$, as $\vartheta_i$ is not a function of $\textbf{p}_i$. Therefore,
\begin{align}
	\frac{\partial}{\partial t} \int_V  \rho  dV = - \displaystyle\sum_{i = 1 }^{N} \bigg\langle \textbf{p}_{i}  \cdot \frac{\partial \vartheta_i }{\partial \textbf{r}}     ; \textit{f}  \bigg\rangle,
\label{EqIK22}
\end{align}
where the equality, $\partial \vartheta_i /\partial \textbf{r}_i  = -\partial \vartheta_i/\partial \textbf{r} $ has been used.  From the continuum mass conservation given in Eq. (\ref{BofmEqn2}), 
the macroscopic and mesoscopic fluxes over the surfaces can be equated, 
\begin{align}
	\displaystyle\sum_{faces}^{6} \int_{S_{f}} \rho \boldsymbol{u} \cdot d\textbf{S}_{f} = \displaystyle\sum_{i = 1 }^{N} \bigg\langle \textbf{p}_{i}  \cdot d\textbf{S}_i    ; \textit{f}  \bigg\rangle.
\label{C2M_mass}
\end{align}
The mesoscopic equation for evolution of mass in a control volume is given by,
\begin{align}
\frac{\partial}{\partial t} \displaystyle\sum_{i=1}^{N}  \bigg\langle m_i \vartheta_i ; \textit{f} \bigg\rangle 
= - \displaystyle\sum_{i = 1 }^{N} \bigg\langle \textbf{p}_{i}  \cdot d\textbf{S}_i    ; \textit{f}  \bigg\rangle.
\label{CVmass_eqn}
\end{align}
Appendix \ref{sec:limitV0IK} shows that the surface mass flux yields the \citet{Irving_Kirkwood} expression for divergence as the CV tends to a point (i.e.\ $V \rightarrow 0$), in analogy to \eq{definition_of_grad}.

\abovesubsectionspace
\subsection{Momentum Balance for a Molecular CV}
\belowsubsectionspace
\label{sec:ddtmomentum}

In this section, a mesoscopic expression for time evolution of momentum within a CV is derived.
The starting point is to integrate the momentum at a point, given in Eq.\ (\ref{Momdensity}), over the CV,
\begin{align}
	\int_V \rho(\textbf{r},t) \boldsymbol{u}(\textbf{r},t) dV 
		=\displaystyle\sum_{i=1}^{N} \bigg\langle \textbf{p}_i  \vartheta_i ; \textit{f} \bigg\rangle.
\label{CVMomdensity}
\end{align}
Following a similar procedure to that in section \ref{sec:ddtmass}, the formula (\ref{EqIK18}) is used to obtain the time evolution of the momentum within the CV,
\begin{align}
	\frac{\partial}{\partial t}\int_V \rho(\textbf{r},t) \boldsymbol{u}(\textbf{r},t) dV  
		= \frac{\partial}{\partial t} \displaystyle\sum_{i=1}^{N} \bigg\langle \textbf{p}_i  \vartheta_i ; \textit{f} \bigg\rangle		\nonumber \\ 
		= \displaystyle\sum_{i = 1 }^{N} \bigg\langle \underbrace{\frac{\textbf{p}_{i}}{m_i} \cdot \frac{\partial}{\partial \textbf{r}_{i}} \textbf{p}_i \vartheta_i}_{\mathcal{K_T}} 
			+  \underbrace{ \textbf{F}_{i} \cdot \frac{\partial}{\partial \textbf{p}_{i}} \textbf{p}_i \vartheta_i }_\mathcal{C_T} ; \textit{f}  \bigg\rangle,
\label{dMdtmeso}
\end{align}
where the terms $\mathcal{K_T}$ and $\mathcal{C_T}$ are the kinetic and configurational components, respectively. 
The kinetic part is,
\begin{align}
	\mathcal{K_T} 
		= \displaystyle\sum_{i = 1 }^{N} \bigg\langle  \frac{\textbf{p}_i}{m_i} \cdot \frac{\partial}{\partial \textbf{r}_i} \textbf{p}_i \vartheta_i ; \textit{f}  \bigg\rangle 
		= \displaystyle\sum_{i = 1 }^{N} \bigg\langle  \frac{\textbf{p}_i \textbf{p}_i }{m_i} \cdot  \frac{\partial \vartheta_i}{\partial \textbf{r}_i}  ; \textit{f}  \bigg\rangle,
\label{KTeq}
\end{align}
where $\textbf{p}_i\textbf{p}_i$ is the dyadic product. For any surface of the CV, here $x^+$, the molecular flux can be equated to the continuum convection and pressure on that surface,
\begin{align}
	\int_{S_{x}^+} \rho(x^+,y,z,t) \boldsymbol{u}(x^+,y,z,t) u_x(x^+,y,z,t)  dydz		\nonumber \\
	+ \int_{S_{x}^+} \textbf{K}_x^{+} dydz = \displaystyle\sum_{i = 1 }^{N} \bigg\langle \frac{\textbf{p}_i p_{ix}  }{m_i} dS_{xi}^+ ; \textit{f}  \bigg\rangle, \nonumber
\end{align}
where $\textbf{K}_x^{+}$ is the kinetic part of the pressure tensor due to molecular transgressions across the $x^+$ CV surface. The average molecular flux across the surface is then,
 \begin{align}
\{\rho \boldsymbol{u} u_x\}^+ + \textbf{K}_x^{+} = \frac{1}{\Delta A_x^+} \displaystyle\sum_{i = 1 }^{N} \bigg\langle \frac{\textbf{p}_i p_{ix}  }{m_i} dS_{xi}^+ ; \textit{f}  \bigg\rangle,
\label{MOP_kinetic}
 \end{align}
where the continuum expression $\{\rho \boldsymbol{u} u_x\}^+$ is the average flux through a flat region in space with area $\Delta A_x^+=\Delta y \Delta z$. 
This kinetic component of the pressure tensor is discussed further in Section \ref{sec:pressure}.

\begin{figure}[H]
	\includegraphics[width=0.48\textwidth]{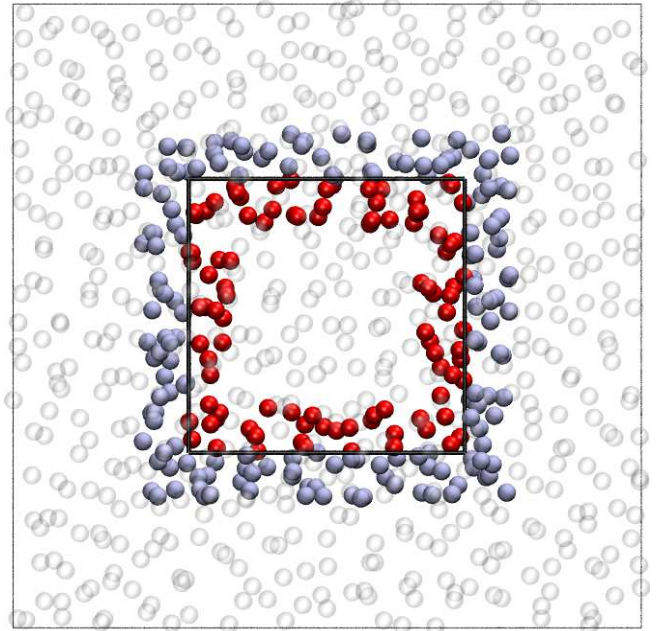}
	\caption{\label{fig:interactions} 
	(Color online) A section through the CV to illustrate the role of $\vartheta_{ij}$ in selecting only the $i$ and $j$ interactions that cross the bounding surface of the control volume. 
	Due to the limited range of interactions, only the forces between the internal (red) molecules $i$ and external  (blue) molecules $j$ near the surfaces are included.}
\figspace
\end{figure}

The configurational term of Eq.\ (\ref{dMdtmeso}) is,
\begin{align}
	\mathcal{C_T} = \displaystyle\sum_{i = 1 }^{N} \bigg\langle \textbf{F}_{i} \cdot \frac{\partial}{\partial \textbf{p}_{i}}  \textbf{p}_i \vartheta_i ; \textit{f}  \bigg\rangle 
				= \displaystyle\sum_{i = 1 }^{N} \bigg\langle \textbf{F}_{i} \vartheta_i ; \textit{f}  \bigg\rangle,
\label{EqIK23}
\end{align}
where the total force $\textbf{F}_i$ on particle $i$ is the sum of pairwise-additive interactions with potential $\phi_{ij}$, and from an external potential $\psi_i$.
\begin{align}
	\vartheta_i \textbf{F}_i  = - \vartheta_i \frac{\partial }{\partial \textbf{r}_{i}} \left( \displaystyle\sum_{\substack{j \neq i }}^{N} \phi_{ij} + \psi_i  \right) . \nonumber
\end{align}
It is commonly assumed that the potential energy of an interatomic interaction, $\phi_{ij}$, can be divided equally between the two interacting molecules, $i$ and $j$, such that,
\begin{align}
	\displaystyle\sum_{i,j}^{N} \vartheta_i \frac{\partial \phi_{ij} }{\partial \textbf{r}_{i}}  
		= \frac{1}{2} \displaystyle\sum_{i,j}^{N} \left[ \vartheta_i \frac{\partial \phi_{ij} }{\partial \textbf{r}_{i}} + \vartheta_j \frac{\partial \phi_{ji} }{\partial \textbf{r}_{j}} \right],
\label{EqIK25}
\end{align}
where the notation $\sum_{i,j}^{N} = \sum_{i=1}^{N} \sum_{j \ne i}^{N}$ has been introduced for conciseness. 
Therefore, the configurational term can be expressed as,
\begin{align}
	\mathcal{C_T} = 
		 \frac{1}{2} \displaystyle\sum_{i,j}^{N}  \bigg\langle \textbf{f}_{ij}  \vartheta_{ij} ; \textit{f}  \bigg\rangle 
		 + \displaystyle\sum_{i=1}^{N} \bigg\langle \textbf{f}_{i_{\textnormal{ext}}} \vartheta_i ; \textit{f}  \bigg\rangle,
\label{CTeq}
\end{align}
where $f_{ij} = -\partial \phi_{ij} / \partial \textbf{r}_{i} = \partial \phi_{ji} /\partial \textbf{r}_{j} $ and $\textbf{f}_{i_{\textnormal{ext}}} = -\partial \psi_{i} /\partial \textbf{r}_{i}$.
The notation, $\vartheta_{ij} \define \vartheta_i - \vartheta_j $, is introduced, which is non-zero only when the force acts over the surface of the CV, as illustrated in Fig.\ \ref{fig:interactions}.

Substituting the kinetic ($K_T$) and configurational ($C_T$) terms, from Eqs.\ (\ref{KTeq}) and (\ref{CTeq}) into Eq.\ (\ref{dMdtmeso}), the time evolution of momentum within the CV at the mesoscopic scale is,
\begin{align}
	\frac{\partial}{\partial t} \displaystyle\sum_{i=1}^{N} \bigg\langle \textbf{p}_i  \vartheta_i ; \textit{f} \bigg\rangle
	\nonumber
	= - \displaystyle\sum_{i = 1}^{N} \bigg\langle  \frac{\textbf{p}_i \textbf{p}_i }{m_i} \cdot  d\textbf{S}_i ; \textit{f}  \bigg\rangle
	\nonumber \\
	+  \frac{1}{2} \displaystyle\sum_{i,j}^{N} \bigg\langle \textbf{f}_{ij} \vartheta_{ij}     ; \textit{f}  \bigg\rangle 
	+ \displaystyle\sum_{i = 1}^{N} \bigg\langle \textbf{f}_{i_{\textnormal{ext}}}  \vartheta_i ; \textit{f}  \bigg\rangle.
\label{meso_timeeqevo}
\end{align}
Equations (\ref{CVmass_eqn}) and (\ref{meso_timeeqevo}) describe the evolution of mass and momentum respectively within a CV averaged over an ensemble of representative molecular systems.  As proposed by \citet{Evans_Morris}, it is possible to develop microscopic evolution equations that do not require ensemble averaging.  Hence, the equivalents of Eqs.\ (\ref{CVmass_eqn}) and (\ref{meso_timeeqevo}) are derived for a single trajectory through phase space in section \ref{sec:micro}, integrated in time in section \ref{sec:timeintmicro} and tested numerically using molecular dynamics simulation in section \ref{sec:MD_results_discussion}. 

The link between the macroscopic and mesoscopic treatments is given by equating their respective momentum Eqs. (\ref{BofMEqn2}) and (\ref{meso_timeeqevo}),
 \begin{align}
	- \oint_S \rho \boldsymbol{u} \boldsymbol{u} \cdot d\textbf{S} +  \textbf{F}_{\textnormal{surface}}   + \textbf{F}_{\textnormal{body}}  
	\nonumber \\ 
	= -\displaystyle\sum_{i = 1}^{N} \bigg\langle  \frac{\textbf{p}_i \textbf{p}_i }{m_i} \cdot  d\textbf{S}_i ; \textit{f}  \bigg\rangle 
	\nonumber \\ 
	 +   \frac{1}{2} \displaystyle\sum_{i,j}^{N} \bigg\langle \textbf{f}_{ij} \vartheta_{ij}     ; \textit{f}  \bigg\rangle
	 + \displaystyle\sum_{i = 1}^{N} \bigg\langle \textbf{f}_{i_{\textnormal{ext}}}  \vartheta_i ; \textit{f}  \bigg\rangle . 
\label{macro2mesomom}
\end{align}
As can be seen, each term in the continuum evolution of momentum has an equivalent term in the mesoscopic formulation. 

The continuum momentum Eq. (\ref{BofMEqn2}) can be expressed in terms of the divergence of the pressure tensor, $\boldsymbol{\Pi}$, in the control volume from,
\begin{subequations}
  \begin{align}
	\frac{\partial}{\partial t}\int_V \rho \boldsymbol{u} dV  = -\oint_S \left[ \rho \boldsymbol{u} \boldsymbol{u} + \boldsymbol{\Pi} \right] \cdot  d\textbf{S} + \textbf{F}_{\textnormal{body}} 
	\label{macrodivergence2mesomoma} \\
	= -\int_V \frac{\partial }{\partial \textbf{r}} \cdot \left[ \rho \boldsymbol{u} \boldsymbol{u} + \boldsymbol{\Pi} \right]  dV + \textbf{F}_{\textnormal{body}}.
	\label{macrodivergence2mesomomb}
  \end{align} 
\end{subequations}
In the following subsection, the right hand side of Eq.\ (\ref{macro2mesomom}) is recast first in divergence form as in Eq.\ (\ref{macrodivergence2mesomomb}), and then in terms of surface pressures as in Eq.\ (\ref{macrodivergence2mesomoma}).

\abovesubsectionspace
\subsection{The Pressure Tensor}

\label{sec:pressure}
\vspace*{-0.1truein}
The average molecular pressure tensor ascribed to a control volume is conveniently expressed in terms of the 
\CV \ function. This is shown \emph{inter alia} to lead to a number of literature definitions of the local stress tensor.  
In the first part of this section, the techniques of \citet{Irving_Kirkwood} are used to express the divergence of the stress 
(as with the right hand side of Eq.\ (\ref{macrodivergence2mesomomb})) in terms of intermolecular force. 
Secondly, the CV pressure tensor is related to the Volume Average (VA) formula (\citep{Lutsko, Cormier_et_al}) and, by consideration of the interactions across the surfaces, to the Method Of Planes (MOP) \citep{Todd_et_al_95, Han_Lee}.
Finally, the molecular CV Eq. (\ref{meso_timeeqevo}) is written in analogous form to the macroscopic Eq. (\ref{macrodivergence2mesomoma}).

The pressure tensor, $ \boldsymbol{\Pi}$, can be decomposed into a kinetic $\boldsymbol{\kappa}$ term, and a configurational stress $\boldsymbol{\sigma}$.  In keeping with the engineering literature, the stress and pressure tensors have opposite signs,
\begin{align}
	\boldsymbol{\Pi} = \boldsymbol{\kappa} - \boldsymbol{\sigma}.
\label{stressdecomp}
\end{align}
The separation into kinetic and configurational parts is made to accommodate the debate concerning the inclusion of kinetic terms in the molecular stress \citep{Zhou, subramaniyan_sun, Hoover_et_al}.

In order to avoid confusion, the stress, $\boldsymbol{\sigma}$, is herein defined to be due to the forces  only (surface tractions).
This, combined with the kinetic pressure term $\boldsymbol{\kappa}$, yields the total pressure tensor $\boldsymbol{\Pi}$ first introduced in \eq{Ftostress}.

\vspace{-0.1in}
\subsubsection{Irving Kirkwood Pressure Tensor} 
\vspace{-0.1in}
The virial expression for the stress cannot be applied locally as it is only valid for a homogeneous system, \citep{Tsai}. The \citet{Irving_Kirkwood} technique for evaluating the non-equilibrium, locally-defined stress resolves this issue, and is herein extended to a CV. To obtain the stress, $\sigma$, the intermolecular force term of Eq.\ (\ref{macro2mesomom}) is defined to be equal to the divergence of stress, 
\begin{align}
	\int_V \frac{\partial}{\partial \textbf{r}} \cdot \boldsymbol{\sigma} dV 
		\define \frac{1}{2} \displaystyle\sum_{i,j}^{N} \bigg\langle \textbf{f}_{ij} \vartheta_{ij} ; \textit{f}  \bigg\rangle \nonumber \\
		=\frac{1}{2}\displaystyle\sum_{i,j}^{N} \int_V \bigg\langle  \textbf{f}_{ij} \left[ \delta(\textbf{r}_i-\textbf{r}) - \delta(\textbf{r}_j-\textbf{r}) \right]  ; \textit{f}  \bigg\rangle dV.
\label{Fsurfpreint}
\end{align}
\citet{Irving_Kirkwood} used a Taylor expansion of the Dirac $\delta$ functions to express 
the pair force contribution in the form of a divergence,
\begin{align}
	\textbf{f}_{ij} \left[ \delta(\textbf{r}_i-\textbf{r}) - \delta(\textbf{r}_j-\textbf{r}) \right] 
		= -\frac{\partial }{\partial \textbf{r}} \cdot \textbf{f}_{ij} \textbf{r}_{ij} O_{ij} \delta(\textbf{r}_i-\textbf{r}), \nonumber
\end{align}
where $\textbf{r}_{ij} = \textbf{r}_i-\textbf{r}_{j}$, and $O_{ij}$ is an operator which acts on the Dirac $\delta$ function,
\begin{align}
	O_{ij} \define \! \left( 1 - \frac{1}{2}\textbf{r}_{ij}\frac{\partial}{\partial \textbf{r}_i} + \ldots - \frac{1}{n!}\left(\textbf{r}_{ij}\frac{\partial}{\partial \textbf{r}_i} \right)^{n-1} \!\!\!\!\! + \ldots \right). 
\label{IKOij}
\end{align}
Equation (\ref{Fsurfpreint}) can therefore be rewritten,
\begin{align}
	\int_V \frac{\partial }{\partial \textbf{r}} \cdot \boldsymbol{\sigma}  dV  
		= -\frac{1}{2} \displaystyle\sum_{i,j}^{N}  \int_V \bigg\langle \frac{\partial }{\partial \textbf{r}}  \cdot \textbf{f}_{ij} \textbf{r}_{ij} \;\;\;\;\;\;\;\;\;\;
	\nonumber \\
	O_{ij} \delta(\textbf{r}_i-\textbf{r}) ; \textit{f} \bigg\rangle  dV.
\label{stress_equality}
\end{align}
The Taylor expansion in Dirac $\delta$ functions is not straightforward to evaluate. This operation can be bypassed by integrating  the position of the molecule $i$ over phase space \citep{Noll}, or by replacing the Dirac $\delta$ with a similar but finite-valued function of compact support \citep{Hardy, Murdoch, Murdoch_2010, Admal_Tadmor}. In the current treatment, the \CV \ function, $\vartheta$, is used, which is advantageous because it explicitly defines both the extent of the CV and its surface fluxes.  
The pressure tensor can be written in terms of the \CV \ function by exploiting the following identities (see Appendix of Ref. \citep{Irving_Kirkwood}),
\begin{align}
	O_{ij} \delta(\textbf{r}_i-\textbf{r}) = \int\limits_{0}^{1}  \delta(\textbf{r} - \textbf{r}_i + s \textbf{r}_{ij}) ds,
\label{intritojds}
\end{align}
Equation\ (\ref{stress_equality}) can therefore be written as,
\begin{align}
	 \int_V \frac{\partial }{\partial \textbf{r}} \cdot \boldsymbol{\sigma}  dV 
	 	=  -\int_{V}  \frac{1}{2}\displaystyle\sum_{i,j}^{N} \bigg\langle \frac{\partial }{\partial \textbf{r}} \cdot \textbf{f}_{ij} \textbf{r}_{ij}  
\nonumber \\
\times  \int\limits_{0}^{1}    \delta(\textbf{r} - \textbf{r}_i + s \textbf{r}_{ij}) ds ; \textit{f}  \bigg\rangle  dV.
\label{divofstress}
 \end{align}
Equation \eq{divofstress} leads to the VA and MOP definitions of the pressure tensor.

\vspace{-0.1in}
\subsubsection{VA Pressure Tensor}
\begin{figure}
\includegraphics[width=0.48\textwidth]{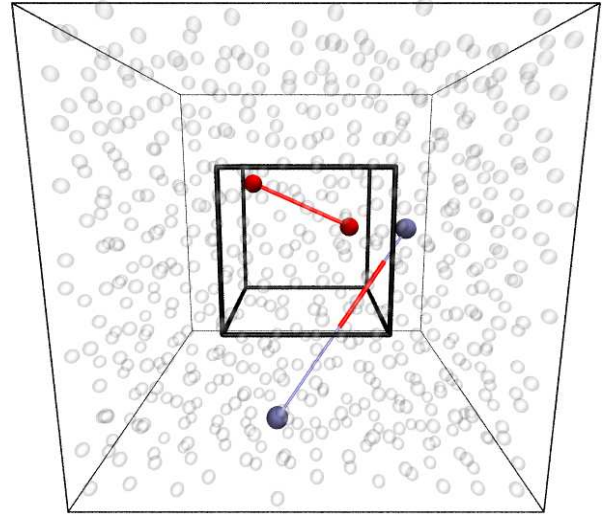}
\caption{\label{fig:VA} (Color online) A plot of the interaction length given by the integral of the selecting function $\vartheta_s$ defined in Eq.\ (\ref{CVfunction_s}) along the line between $r_i$ and $r_j$. 
The cases shown are for two molecules which are \textit{a)} 
both inside the volume ($l_{ij} = 1$) and \textit{b)} 
both outside the volume with an interaction crossing the volume, where $l_{ij}$ is the 
fraction of the total length between $i$ and $j$ inside the volume. The line is thin (blue) outside and thicker (red) inside the volume.}
\figspace
\end{figure}

\vspace{-0.1in}
definition of the stress tensor of \citet{Lutsko} and \citet{Cormier_et_al} can be obtained by rewriting \eq{divofstress} as,
\begin{align}
	\frac{\partial }{\partial \textbf{r}} \cdot  \int_V \boldsymbol{\sigma}  dV 
	 	=  - \frac{\partial }{\partial \textbf{r}} \cdot \int_{V}  \frac{1}{2}\displaystyle\sum_{i,j}^{N} \bigg\langle  \textbf{f}_{ij} \textbf{r}_{ij}  
\nonumber \\
\times  \int\limits_{0}^{1}    \delta(\textbf{r} - \textbf{r}_i + s \textbf{r}_{ij}) ds ; \textit{f}  \bigg\rangle  dV.
\label{divofintofstress}
 \end{align}
Equating the expressions inside the divergence on both sides of Eq.\ (\ref{divofintofstress}), \footnote{The resulting equality satisfies Eq.\ (\ref{divofintofstress}) and both sides are equal to within an arbitrary constant (related to choosing the gauge).}, and assuming the stress is constant within an arbitrary local volume, $\Delta V$, gives an expression for the VA stress,
\begin{align}
	\StressVAcauchy  = -\frac{1}{2 \Delta V} \! \int_V \displaystyle\sum_{i,j}^{N} \bigg\langle \textbf{f}_{ij} \textbf{r}_{ij} \!  \int\limits_{0}^{1} \!\!    \delta(\textbf{r} - \textbf{r}_i + s \textbf{r}_{ij}) ds ; \textit{f}  \bigg\rangle dV.
\label{constant_volume_stress}
\end{align}
Swapping the order of integration and evaluating the integral of the Dirac $\delta$ function over $\Delta V$ gives a different form of the \CV \ function, $\vartheta_s$,
\begin{align}
	\vartheta_s \define \int_V \delta(\textbf{r} - \textbf{r}_i + s \textbf{r}_{ij}) dV =  \;  \;  \;
	\nonumber \\
	\left[ H(x^+ - x_i + s x_{ij}) - H(x^- - x_i + s x_{ij}) \right]  \; \;
	\nonumber \\
	\times\left[ H(y^+ - y_i + s y_{ij}) \! \; - \, H(y^- - y_i + s y_{ij}) \right] \; \;
	\nonumber \\
	\times\left[ H(z^+ - z_i + s z_{ij}) \; - \; H(z^- - z_i + s z_{ij}) \right],
\label{CVfunction_s}
\end{align}
which is non-zero if a point on the line between the two molecules, $\textbf{r}_i - s \textbf{r}_{ij}$, is inside the cubic region (c.f. $\textbf{r}_i$ with $\vartheta_i$). Substituting the definition, $\vartheta_s$ (Eq.\ \ref{CVfunction_s}), into Eq.\ (\ref{constant_volume_stress}) gives,
\begin{align}
	\StressVAcauchy = -\frac{1}{2\Delta V} \displaystyle\sum_{i,j}^{N} \bigg\langle \textbf{f}_{ij} \textbf{r}_{ij}  l_{ij}  ; \textit{f}  \bigg\rangle, 
\label{VA}
\end{align}
where $l_{ij}$ is the integral from $r_i$ ($s=0)$ to $r_j$ ($s=1$) of the $\vartheta_s$ function,
\begin{align}
l_{ij} \define \int_{0}^{1}  \vartheta_s ds. \nonumber
 \end{align}
Therefore, $l_{ij}$ is the fraction of interaction length between $i$ and $j$ 
which lies within the CV, as illustrated in Fig.\ \ref{fig:VA}.
The definition of the configurational stress in Eq.\ (\ref{VA}) is the same as in the work of \citet{Lutsko} and \citet{Cormier_et_al}. The microscopic divergence theorem given in Appendix \ref{sec:divergence_ReynoldsTT} can be applied to obtain the volume averaged kinetic component 
of the pressure tensor, $\mathcal{K_T}$, in Eq.\ (\ref{KTeq}), 
\begin{align}
	\displaystyle\sum_{i = 1}^{N} \bigg\langle  \frac{\textbf{p}_i \textbf{p}_i }{m_i} \cdot d\textbf{S}_i ; \textit{f}  \bigg\rangle 
	= \frac{\partial}{\partial \textbf{r}}  \cdot \displaystyle\sum_{i = 1}^{N} \overbrace{\bigg\langle  \frac{\textbf{p}_i \textbf{p}_i }{m_i}  \vartheta_i  ; \textit{f}  \bigg\rangle}^{{\rhouuVA + \PressureVAkinetic}  }. \nonumber
\end{align}
Note that the expression inside the divergence includes both the advection, $\rhouuVA$, and kinetic components of the pressure tensor. 
The VA form \citep{Cormier_et_al} is obtained by combining the above expression with the configurational stress $\StressVAcauchy$,
\begin{align}
	\rhouuVA + {\PressureVAkinetic} - \StressVAcauchy 
	= \rhouuVA + \PressureVA \;\;\;\; \;\;\;\;\;\;\;\; \;\;\;\;  \nonumber \\
	\;\;\;\;\;\;\;\; \;\;\;\;  = \frac{1}{\Delta V} \displaystyle\sum_{i = 1 }^{N} \bigg\langle  \frac{\textbf{p}_i \textbf{p}_{i}}{m_i}  \vartheta_i + \frac{1}{2}\displaystyle\sum_{i,j}^{N}    \textbf{f}_{ij} \textbf{r}_{ij} l_{ij} ; \textit{f} \bigg\rangle. 
\label{VA_pressure}
\end{align}
In contrast to the work of \citet{Cormier_et_al}, the advection term in the above expression is explicitly identified, in order to be compatible with the right hand side of Eq. (\ref{macrodivergence2mesomomb}) and definition of the pressure tensor, $\boldsymbol{\Pi}$.

\vspace{-0.2in}
\subsubsection{MOP Pressure Tensor}
\vspace{-0.1in}
The stress in the CV can also be related to the tractions over each surface. In analogy to prior use of the molecular \CV \ function, $\vartheta_i$, to evaluate the flux, the stress \CV \ function, $\vartheta_s$, can be differentiated to give the tractions over each surface. These surface tractions are the ones used in the formal definition of the continuum Cauchy stress tensor. The surface traction (i.e., force per unit area) and the kinetic pressure on a surface combined give the MOP expression for the pressure tensor \citep{Todd_et_al_95}.

In the context of the CV, the forces and fluxes on the six bounding surfaces are required to obtain the pressure inside the CV.
It is herein shown that each face takes the form of the \citet{Han_Lee} localization of the MOP pressure components. 
The divergence theorem is used to express the left hand side of Eq.\ (\ref{divofstress}) in terms of stress across the six faces of the cube. The mesoscopic right hand side of Eq.\ (\ref{divofstress}) can also be expressed as surface stresses by starting with the \CV \ function $\vartheta_s$, 
 \begin{align}
	\displaystyle\sum_{faces}^{6} \int_{S_{f}} \boldsymbol{\sigma} \cdot d\textbf{S}_{f} =  - \frac{1}{2}\displaystyle\sum_{i,j}^{N} \bigg\langle   	\textbf{f}_{ij} \textbf{r}_{ij} \cdot  \int\limits_{0}^{1} \frac{\partial  \vartheta_s }{\partial \textbf{r}}  ds ; \textit{f}  \bigg\rangle. \nonumber
 \end{align}
The procedure for taking the derivative of $\vartheta_s$ with respect to $\textbf{r}$ 
and integrating over the volume is given in Appendix \ref{sec:lineplane}. 
The result is an expression for the force on the CV rewritten as the force over each surface of the CV. For the $x^+$ face, for example, this is,
\begin{align}
	\int_{S^+_{x}} \boldsymbol{\sigma} \cdot d\textbf{S}_{S^+_{x}}  
		 =- \frac{1}{4} \displaystyle\sum_{i,j}^{N} \bigg\langle  \textbf{f}_{ij} \left[\; sgn(x^+ - x_j) \;\;\;\;\;\;\;\;\;\;\;\;\;\;
	\nonumber \nl 
	- sgn(x^+ - x_i)\right]S_{xij}^+   ; \textit{f}  \bigg\rangle. \nonumber
 \end{align}
The combination of the signum functions and the $S_{xij}^+$ term specifies when the point of intersection of the line between $i$ and $j$ 
is located on the $x^+$ surface of the cube (see Appendix \ref{sec:lineplane}). Corresponding expressions for the $y$ and $z$ faces are defined by $S_{\alpha ij}^\pm$ when $\alpha = \{y,z\}$ respectively. 

The full expression for the MOP pressure tensor, which includes the kinetic part given by Eq.\ (\ref{MOP_kinetic}), is obtained by assuming a uniform pressure over the $x^+$ surface,
\begin{align}
	\int_{S^+_{x}} \boldsymbol{\Pi} \cdot d\textbf{S}_{x}^+ 
		= \left[ \boldsymbol{\kappa} - \boldsymbol{\sigma} \right] \cdot \textbf{n}_x^+ \Delta A_x^+ 
	\nonumber \\
		\define  \left[ \textbf{K}_x^{+} - \boldsymbol{T}_x^{+} \right] \Delta A_x^+ = \textbf{P}_x^{+}  \Delta A_x^+ ,
\end{align}
where $\textbf{n}_x^+$ is a unit vector aligned along the $x$ coordinate axis, $\textbf{n}_x^+ = [+1 , 0 , 0]$; $\boldsymbol{T}_x^{+}$ is the configurational stress (traction) and $\textbf{P}_x^{+}$ the total pressure tensor acting on a plane.
Hence, 

\begin{align}
	\textbf{P}_x^+ = \frac{1}{\Delta A_x^+} \displaystyle
		\sum_{i = 1 }^{N} \bigg\langle \frac{\overline{\textbf{p}}_i \overline{p}_{ix}}{m_i} \delta(x_i-x^+) S_{xi}^+ ; \textit{f}  \bigg\rangle \;\;\;\;\;\;\;\;\;\;\;\;\;\;\;\;\;\;\;\;\;\;\;\;\;\;\;\;\;\;
	\nonumber \\ 
	+ \frac{1}{4\Delta A_x^+} \displaystyle\sum_{i,j}^{N} \bigg\langle \textbf{f}_{ij} \left[\! sgn(x^+ \! - \! x_j) 
	- sgn(x^+ \! - \! x_i) \!\right] \! S_{xij}^+   ; \textit{f}  \bigg\rangle, \;\;\;\;\;
\label{CV_pressure}
\end{align}
where the peculiar momentum, $\overline{\textbf{p}}_i$ has been used as in \citet{Todd_et_al_95}. 
If the $x^+$ surface area covers the entire domain ($S_{xij}^+=1$ in \eq{CV_pressure}), the MOP formulation of the pressure is recovered \citep{Todd_et_al_95}. 

\begin{figure}
\includegraphics[width=0.48\textwidth]{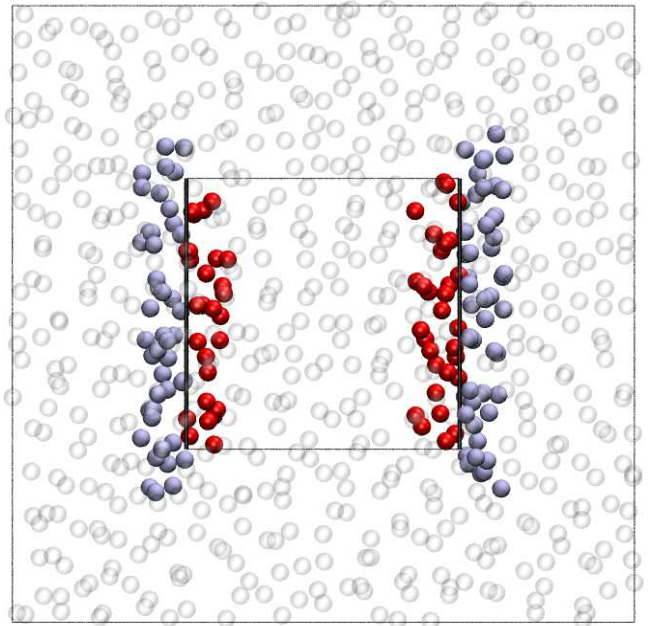}
\caption{\label{fig:plane_interactions} 
(Color online) Representation of those molecules selected through $dS_{xij}$ in 
Eq.\ (\ref{micro_face_stress}) with molecules $i$ on the side of the surface inside the CV
(red) and molecules $j$ on the outside (blue). The CV is the inner square on the figure.}
\figspace
\end{figure}

The extent of the surface is defined through $S_{xij}^+$, in Eq.\ (\ref{CV_pressure}) 
which is the localized form of the pressure tensor considered by \citet{Han_Lee} applied to the six cubic faces. 
For a cube in space, each face has three components of stress, which results in $18$ independent components over the total control surface. The quantity,
 \begin{align}
dS_{\alpha ij} \define \frac{1}{2}\left[sgn(r^+_\alpha - r_{\alpha j}) - sgn(r^+_\alpha - r_{\alpha i})\right] S_{\alpha ij}^+ \;
\nonumber \\
 - \frac{1}{2}\left[sgn(r^-_\alpha - r_{\alpha j}) - sgn(r^-_\alpha - r_{\alpha i})\right] S_{\alpha ij}^- , \nonumber
 \end{align}
selects the force contributions across the two opposite faces; similar notation to the surface molecular flux, 
$d\textbf{S}_{ij} = d\textbf{S}_{ij}^+ - d\textbf{S}_{ij}^-$ (c.f.\ \!\!\!\! \eq{dSi}), is used. The case of the two $x$ planes located on opposite sides of the cube is illustrated in Fig.\ \ref{fig:plane_interactions}. 

Taking all surfaces of the cube into account yields the final form,
\begin{align}
	\displaystyle\sum_{faces}^{6} \int_{S_{f}} \boldsymbol{\sigma} \cdot d\textbf{S}_{f} 
		= -\frac{1}{2} \displaystyle\sum_{i,j}^{N} \bigg\langle \textbf{f}_{ij} \sum_{\alpha=1}^3 dS_{\alpha ij} ; \textit{f} \bigg\rangle
\nonumber \\
= -\frac{1}{2} \displaystyle\sum_{i,j}^{N} \bigg\langle\textbf{f}_{ij} \tilde{\textbf{n}} \cdot d\textbf{S}_{ ij} ; \textit{f} \bigg\rangle
\nonumber \\
= \frac{1}{2} \displaystyle\sum_{i,j}^{N} \bigg\langle \boldsymbol{\varsigma}_{ij}\cdot d\textbf{S}_{ ij} ; \textit{f} \bigg\rangle.
\label{micro_face_stress} 
\end{align}
The vector $\tilde{\textbf{n}}$, obtained in Appendix \ref{sec:lineplane}, is unity in each direction. The tensor $\boldsymbol{\varsigma}_{ij}$ is defined, for notational convenience, to be the outer product of the intermolecular forces with $\tilde{\textbf{n}}$,
\begin{align}
\boldsymbol{\varsigma}_{ij} \define  
-\textbf{f}_{ij} \tilde{\textbf{n}} =
-\textbf{f}_{ij} \begin{bmatrix}
  1 & 1 & 1 
 \end{bmatrix} =
-\begin{bmatrix}
  f_{xij} & f_{xij} & f_{xij} \\
  f_{yij} & f_{yij} & f_{yij} \\
  f_{zij} & f_{zij} & f_{zij} 
 \end{bmatrix}  \nonumber.
\end{align}
In this form, the $\vartheta_{ij}$ function for all interactions over the cube's surface is expressed as the sum of six selection functions for each of the six faces, i.e. $\vartheta_{ij} = - \sum_{\alpha=1}^3 dS_{\alpha ij}$. 

\vspace{-0.1in}
\subsubsection{Relationship to the continuum}
\vspace{-0.1in}
The forces per unit area, or 'tractions', acting over each face of the CV, are used in the definition of the Cauchy stress tensor at the continuum level. 
For the $x^+$ surface, the traction vector is the sum of all forces acting over the surface,
\begin{align}
	\textbf{T}_x^+ 	= - \frac{1}{4\Delta A_x^+} \! \displaystyle\sum_{i,j}^{N} \bigg\langle \textbf{f}_{ij} \! \left[sgn(x^+ - x_j) \;\;\;\;\;\;\;\;\;\;\;\;\;
	\nonumber \nl 
	- sgn(x^+ - x_i)\right] \! S_{xij}^+ ; \textit{f} \bigg\rangle, 
\label{deftraction}
\end{align}
which satisfies the definition,
\begin{align}
	\textbf{T}_x^\pm  = \boldsymbol{\sigma}  \cdot \textbf{n}_x^\pm, \nonumber
\end{align}
of the Cauchy traction \citep{Nemat-Nasser}. A similar relationship can be written for both the kinetic and total pressures,
\begin{subequations}
 \begin{align}
	\boldsymbol{K}_x^{\pm}  = \boldsymbol{\kappa}  \cdot \textbf{n}_x^\pm, \nonumber \\
	\textbf{P}_x^{\pm}  = \boldsymbol{\Pi}  \cdot \textbf{n}_x^\pm, \nonumber 
 \end{align}
\end{subequations}
where $\textbf{n}_x^\pm$ is a unit vector, $\textbf{n}_x^\pm = [\pm 1 \; \; 0 \; \; 0]^T$. 

The time evolution of the molecular momentum within a CV (\eq{meso_timeeqevo}), can be expressed in a similar form to the Navier-Stokes equations of continuum fluid mechanics. Dividing both sides of \eq{meso_timeeqevo} by the volume, the following form can be obtained;  note that this step requires  Eqs. \eq{MOP_kinetic}, \eq{CV_pressure} and \eq{deftraction}: 
\begin{align}
	\frac{1}{\Delta V}\frac{\partial}{\partial t} \displaystyle\sum_{i=1}^{N} \bigg\langle p_{\alpha i}  \vartheta_i ; \textit{f} \bigg\rangle 
		+ \frac{\{\rho u_\alpha u_\beta\}^+ - \{\rho u_\alpha u_\beta\}^-}{\Delta r_\beta}  = 
	\nonumber \\
	- \frac{K_{\alpha \beta}^{+} - K_{\alpha \beta}^{-}}{\Delta r_\beta}  
		+  \frac{T_{\alpha \beta}^+ - T_{\alpha \beta}^-}{\Delta r_\beta} 
		+  \frac{1}{\Delta V} \displaystyle\sum_{i = 1}^{N} \bigg\langle f_{\alpha i_{\textnormal{ext}}}  \vartheta_i ; \textit{f}  \bigg\rangle, 
\label{mesoevo_traction}
\end{align}
where index notation has been used (e.g. $\textbf{T}_{x}^\pm = T_{\alpha x}^\pm$) with the Einstein summation convention. 

In the limit of zero volume, each expression would be similar to a term in the differential continuum equations (although the pressure term would be the divergence of a tensor and not the gradient of a scalar field as is common in fluid mechanics). 
The Cauchy stress tensor, $\boldsymbol{\sigma}$, is defined in the limit that the cube's volume tends to zero, so that $\textbf{T}^+ $ and $\textbf{T}^-$ are related by an infinitesimal difference. This is used in continuum mechanics to define the unique nine component Cauchy stress tensor, $d\boldsymbol{\sigma}/dx \equiv \lim_{\Delta x \rightarrow 0} [\textbf{T}^+ + \textbf{T}^-]/\Delta x$. This limit is shown in Appendix \ref{sec:limitV0IK} to yield the \citet{Irving_Kirkwood} stress in terms of the Taylor expansion in Dirac $\delta$ functions.

Rather than defining the stress at a point, the tractions can be compared to their continuum counterparts in a fluid mechanics control volume or a solid mechanics Finite Elements (FE) method. Computational Fluid Dynamics (CFD) is commonly formulated using CV and in discrete simulations, Finite Volume \citep{Hirsch}. Surface forces are ideal for coupling schemes between MD and CFD. Building on the pioneering work of \citet{OConnell_Thompson}, there are many MD to CFD coupling schemes -- see the review paper by \citet{Mohamed_Mohamad}. More recent developments for coupling to fluctuating hydrodynamics are covered in a review by \citet{Delgado_Buscalioni_2012}. A discussion of coupling schemes is outside the scope of this work, however finite volume algorithms have been used extensively in coupling methods \citep{Nie_et_al, Werder_et_al, Delgado-Buscalioni_Coveney_04,DeFabritiis_et_al_06, Delgado-Buscalioni_DeFabritiis} together with equivalent control volumes defined in the molecular region.  An advantage of the herein proposed molecular CV approach is that it ensures conservation laws are satisfied when exchanging fluxes over cell surfaces --- an important requirement for accurate unsteady coupled simulations as outlined in the finite volume coupling of \citet{Delgado-Buscalioni_Coveney_04}. For solid coupling schemes, \citep{Curtin_Miller}, the principle of virtual work can be used with tractions on the element corners (the MD CV) to give the state of stress in the element \citep{Zienkiewicz},
 \begin{align}
\int_V \boldsymbol{\sigma} \cdot \boldsymbol{\nabla}N_a dV = \oint_S N_a \textbf{T} dS,
\label{FEA}
 \end{align}
where $N_a$ is a linear shape function which allows stress to be defined as a continuous function of position. It will be demonstrated numerically in the next section, \ref{sec:implement}, that the CV formulation is exactly conservative: the surface tractions and fluxes entirely define the stress within the volume. The tractions and stress in Eq.\ (\ref{FEA}) are connected by the weak formulation and the form of the stress tensor results from the choice of shape function $N_a$. 

\abovesubsectionspace
\subsection{Energy Balance for a Molecular CV}
\belowsubsectionspace
\label{sec:ddtenergy}

In this section, a mesoscopic expression for time evolution of energy within a CV is derived.
As for mass and momentum, the starting point is to integrate the energy at a point, given in Eq.\ (\ref{Energydensity}), over the CV,
\begin{align}
\int_V \rho(\textbf{r},t) \mathcal{E}(\textbf{r},t) dV 
=\displaystyle\sum_{i=1}^{N} \bigg\langle e_i  \vartheta_i ; \textit{f} \bigg\rangle.
\label{CVEnergy}
\end{align}
The time evolution within the CV is given using formula (\ref{EqIK18}),
\begin{align}
\frac{\partial}{\partial t}\int_V \rho(\textbf{r},t) \mathcal{E}(\textbf{r},t) dV  
= \frac{\partial}{\partial t} \displaystyle\sum_{i=1}^{N} \bigg\langle e_i  \vartheta_i ; \textit{f} \bigg\rangle
\nonumber \\ 
= \displaystyle\sum_{i = 1 }^{N} \bigg\langle \frac{\textbf{p}_{i}}{m_i} \cdot \frac{\partial}{\partial \textbf{r}_{i}} e_i \vartheta_i
+  \textbf{F}_{i} \cdot \frac{\partial}{\partial \textbf{p}_{i}} e_i \vartheta_i ; \textit{f}  \bigg\rangle.
\label{dEdtmeso}
\end{align}
Evaluating the derivatives of the energy and \CV ~function results in,
\begin{align}
\frac{\partial}{\partial t} \displaystyle\sum_{i=1}^{N} \bigg\langle e_i  \vartheta_i ; \textit{f} \bigg\rangle		
=-\frac{1}{2} \displaystyle\sum_{i,j}^{N}  \bigg\langle \left[ \frac{\textbf{p}_{i}}{m_i} \cdot \textbf{f}_{ij}+ \frac{\textbf{p}_{j}}{m_i} \cdot \textbf{f}_{ji} \right]  \vartheta_{i} ; \textit{f}  \bigg\rangle  \nonumber \\ 
 - \displaystyle\sum_{i = 1 }^{N} \bigg\langle  e_i \frac{\textbf{p}_{i}}{m_i} \cdot  d\textbf{S}_i -  \textbf{F}_{i} \cdot \frac{\textbf{p}_{i}}{m_i} \vartheta_i ; \textit{f}  \bigg\rangle. \nonumber
\end{align}
Using the definition of $\textbf{F}_i$, Newton's 3rd law and relabelling indices, the intermolecular force terms can be expressed in terms of the interactions over the CV surface, $\vartheta_{ij}$,
 \begin{align}
\frac{\partial}{\partial t} \displaystyle\sum_{i=1}^{N} \bigg\langle e_i  \vartheta_i ; \textit{f} \bigg\rangle		
= - \displaystyle\sum_{i = 1 }^{N} \bigg\langle  e_i \frac{\textbf{p}_{i}}{m_i} \cdot  d\textbf{S}_i ; \textit{f}  \bigg\rangle 
\nonumber \\ 
+ \frac{1}{2} \displaystyle\sum_{i,j}^{N}  \bigg\langle \frac{\textbf{p}_{i}}{m_i} \cdot \textbf{f}_{ij}  \vartheta_{ij} ; \textit{f}  \bigg\rangle  + \displaystyle\sum_{i=1}^{N} \bigg\langle \frac{\textbf{p}_{i}}{m_i} \cdot \textbf{f}_{i_{\textnormal{ext}}} \vartheta_i ; \textit{f}  \bigg\rangle . \nonumber
\end{align}
The right hand side of this equation is equated to the right hand side of the continuum energy Eq.\ \ref{BofenergyEqn},
 \begin{align}
  \overbrace{- \oint_S \rho \mathcal{E} \boldsymbol{u} \cdot d\textbf{S}}^{\text{energy flux}} 
- \overbrace{\oint_S \textbf{q} \cdot d\textbf{S} }^{\text{heat flux}}
- \overbrace{ \oint_S \boldsymbol{\Pi} \cdot \boldsymbol{u} \cdot d\textbf{S}}^{\text{pressure heating}} 
\nonumber \\
=- \displaystyle\sum_{i = 1 }^{N} \bigg\langle  e_i \frac{\textbf{p}_{i}}{m_i} \cdot  d\textbf{S}_i ; \textit{f}  \bigg\rangle 
\nonumber \\
+ \frac{1}{2} \displaystyle\sum_{i,j}^{N}  \bigg\langle \frac{\textbf{p}_{i}}{m_i} \cdot \boldsymbol{\varsigma}_{ij}\cdot d\textbf{S}_{ ij}  ; \textit{f}  \bigg\rangle,
\label{meso_timeEevo}
\end{align}
where the energy due to the external (body) forces is neglected. The $\textbf{f}_{ij}\vartheta_{ij}$ has been re-expressed in terms of surface tractions, $\boldsymbol{\varsigma}_{ij}\cdot d\textbf{S}_{ij}$, using the analysis of the previous section. In its current form, the microscopic equation does not delineate the contribution due to energy flux, heat flux and pressure heating. To achieve this division, the notion of the peculiar momentum at the molecular location, $\boldsymbol{u}(\textbf{r}_i)$ is used together with the velocity at the CV surfaces $\boldsymbol{u}(\textbf{r}^\pm)$, following a similar process to \citet{Evans_Morris}.

\abovesectionspace
\section{Implementation}
\belowsectionspace
\label{sec:implement}

In this section, the CV equation for mass, momentum and energy balance, Eqs.\ (\ref{CVmass_eqn}), (\ref{meso_timeeqevo}) and (\ref{meso_timeEevo}), will be proved to apply and demonstrated numerically for a microscopic 
system undergoing a single trajectory through phase space.

\abovesubsectionspace
\subsection{The Microscopic System}
\belowsubsectionspace
\label{sec:micro}

Consider a single trajectory of a set of molecules through phase space, defined in terms of their time 
dependent coordinates $\textbf{r}_i$ and momentum $\textbf{p}_i$. The \CV \ function depends on molecular coordinates,
the location of the center of the cube, $\textbf{r}$, and its side length, $\Delta \textbf{r}$, 
{\it i.e.,} $\vartheta_i \equiv \vartheta_i (\textbf{r}_i(t),\textbf{r}, \Delta \textbf{r})$. 
The time evolution of the mass within the molecular control volume is given by, 
\begin{align}
 \frac{d}{d t}\displaystyle\sum_{i=1}^{N}  m_i \vartheta_i(\textbf{r}_i(t),\textbf{r}, \Delta \textbf{r})  = \displaystyle\sum_{i = 1}^{N}  m_i \frac{\partial  \vartheta_i}{\partial t} 
\nonumber \\  
 = \displaystyle\sum_{i = 1}^{N}  m_i \frac{d  \textbf{r}_i}{d t} \cdot  \frac{\partial  \vartheta_i}{\partial \textbf{r}_i} = -\displaystyle\sum_{i = 1}^{N}  \textbf{p}_i  \cdot  d\textbf{S}_i, \label{microfluxm}
\end{align}
using, $\textbf{p}_i = m_i d{\textbf{r}}_i/dt$. The time evolution of momentum in the molecular control volume is,
\begin{align}
 \frac{\partial}{\partial t}\displaystyle\sum_{i=1}^{N}  \textbf{p}_i(t) \vartheta_i(\textbf{r}_i(t),\textbf{r}, \Delta \textbf{r}) 
\nonumber \\  
 = \displaystyle\sum_{i = 1}^{N} \left[  \textbf{p}_i   \frac{\partial  \vartheta_i}{\partial t} + \frac{d \textbf{p}_i }{d t}   \vartheta_i  \right]  \;
\nonumber \\  
 =  \displaystyle\sum_{i = 1}^{N} \left[ \textbf{p}_i \frac{d  \textbf{r}_i}{d t} \cdot  \frac{\partial  \vartheta_i}{\partial \textbf{r}_i} +    \frac{d \textbf{p}_i }{d t}   \vartheta_i   \right]. \nonumber
\end{align}
As, $d\textbf{p}_i/dt = \textbf{F}_i $, 
then,
\begin{align}
 \frac{\partial}{\partial t}\displaystyle\sum_{i=1}^{N}  \textbf{p}_i \vartheta_i
 =  \displaystyle\sum_{i = 1}^{N} \left[ - \frac{\textbf{p}_i \textbf{p}_i}{m_i}  \cdot  d\textbf{S}_i + \textbf{F}_i   \vartheta_i  \right] 
\nonumber \\  
 = - \! \displaystyle\sum_{i=1}^{N} \! \frac{\textbf{p}_i \textbf{p}_i}{m_i}  \cdot  d\textbf{S}_i + \frac{1}{2}\displaystyle\sum_{i,j}^{N} \textbf{f}_{ij}  \vartheta_{ij} + \displaystyle\sum_{i=1}^{N} \textbf{f}_{i_{\textnormal{ext}}} \vartheta_i, 
\label{microfluxM}
\end{align}
where the total force on molecule $i$ has been decomposed into surface and `external' or body terms. The time evolution of energy in a molecular control volume is obtained by evaluating,
 \begin{align}
 \frac{\partial}{\partial t}\displaystyle\sum_{i=1}^{N}  e_i \vartheta_i
 =  \displaystyle\sum_{i = 1}^{N} \left[ e_i  \frac{\partial  \vartheta_i}{\partial t} + \frac{\partial e_i}{\partial t}   \vartheta_i  \right] 
\nonumber \\  
 = - \! \displaystyle\sum_{i=1}^{N} \! e_i \frac{\textbf{p}_i}{m_i}  \cdot  d\textbf{S}_i +  \displaystyle\sum_{i=1}^{N} \frac{\dot{\textbf{p}}_i \cdot \textbf{p}_i}{m_i} \vartheta_i  
\nonumber \\  
-  \frac{1}{2}\displaystyle\sum_{i,j}^{N} \left[ \frac{\textbf{p}_{i}}{m_i} \cdot \textbf{f}_{ij}+ \frac{\textbf{p}_{j}}{m_j} \cdot \textbf{f}_{ji} \right]  \vartheta_{i} \nonumber
\end{align}
using, $d\textbf{p}_i/dt = \textbf{F}_i $ and the decomposition of forces. The manipulation proceeds as in the mesoscopic system to yield,
 \begin{align}
 \frac{\partial}{\partial t}\displaystyle\sum_{i=1}^{N}  e_i \vartheta_i
 =  - \! \displaystyle\sum_{i=1}^{N} \! e_i \frac{\textbf{p}_i}{m_i}  \cdot  d\textbf{S}_i
\nonumber \\  
   + \frac{1}{2}\displaystyle\sum_{i,j}^{N} \frac{\textbf{p}_i}{m_i}  \cdot \textbf{f}_{ij}  \vartheta_{ij} + \displaystyle\sum_{i=1}^{N} \frac{\textbf{p}_i}{m_i}  \cdot \textbf{f}_{i_{\textnormal{ext}}} \vartheta_i, 
\label{microfluxE}
\end{align}
The average of many such trajectories defined through Eqs.\  (\ref{microfluxm}), (\ref{microfluxM}) and (\ref{microfluxE})
gives the mesoscopic expressions in Eqs.\ (\ref{CVmass_eqn}), (\ref{meso_timeeqevo}) and (\ref{meso_timeEevo}), respectively. 
In the next subsection, the time integral of the single trajectory is considered.

\abovesubsectionspace
\subsection{Time integration of the microscopic CV equations}
\belowsubsectionspace
\label{sec:timeintmicro}

Integration of 
Eqs. (\ref{microfluxm}), (\ref{microfluxM}) and (\ref{microfluxE}) over the time interval $[0,\tau]$ enables these equations to be usable in a molecular simulation. For the conservation of mass term,
\begin{align}
\displaystyle\sum_{i=1}^{N} m_i \left[ \vartheta_i(\tau) - \vartheta_i(0)  \right] =  - \int\limits_{0}^{\tau} \displaystyle\sum_{i = 1 }^{N} \textbf{p}_{i}  \cdot d\textbf{S}_i dt.
\label{micromassflux}
\end{align}
The surface crossing term, $d\textbf{S}_i$, defined in \eq{flux}, involves a Dirac $\delta$ function and therefore cannot be evaluated directly. Over the time interval $[0,\tau]$, molecule $i$ 
passes through a given $x$ position at times, $t_{xi,k}$, where $k=1,2,...,N_{t_x}$ \citep{Daivis_et_al} .
The positional Dirac $\delta$ can be expressed as, 
\begin{align}
\delta(x_i(t)-x) =  \displaystyle\sum_{k=1}^{N_{t_x}} \frac{\delta(t - t_{xi,k})}{|\dot{x}_i(t_{xi,k})|},
\label{diracfnoft}
\end{align}
where $|\dot{x}_i(t_{xi,k})|$ is the magnitude of the velocity in the $x$ direction at time $t_{xi,k}$.
Equation \eq{diracfnoft} is used to rewrite $d\textbf{S}_i$ in \eq{micromassflux} in the form,
\begin{align}
 d S_{\alpha i ,k} \define \! \left[ sgn(t_{\alpha i,k}^+ \! - \tau) - sgn(t_{\alpha i,k}^+  - \! 0)\right] S_{\alpha i,k}^+(t_{\alpha i,k}^+)  \;
\nonumber \\
- \! \left[ sgn(t_{\alpha i,k}^- \! - \! \tau) - sgn(t_{\alpha i,k}^- - \! 0)\right] S_{\alpha i,k}^-(t_{\alpha i,k}^-),
\label{surfaceflux}
\end{align}
where $\alpha = \{x,y,z\}$, and the fluxes are evaluated at times, $t_{\alpha i,k}^+$
and $t_{\alpha i,k}^-$  for the right and left surfaces of the cube, respectively.
Using the above expression, the time integral in \eq{micromassflux} can be expressed as the sum of all molecule crossings, 
$N_t = N_{t_x} + N_{t_y} + N_{t_z}$ over the cube's faces, 
\begin{align}
\overbrace{\displaystyle\sum_{i=1}^{N} m_i \left[ \vartheta_i(\tau) - \vartheta_i(0)  \right]}^{\text{Accumulation}} 
 = \underbrace{ -\displaystyle\sum_{i = 1 }^{N} \displaystyle\sum_{k=1}^{N_t} m_i \sum_{\alpha=1}^3 \frac{p_{\alpha i}}{|p_{\alpha i}|} dS_{\alpha i ,k}}_{\text{Advection}}.
\label{massfluxMD}
\end{align}
In other words, the mass in a CV at time $t=\tau$ minus its initial value at $t=0$ is the sum of all 
molecules that cross its surfaces during the time interval.

The momentum balance equation \eq{microfluxM}, can also be written in time-integrated form,
\begin{align}
\displaystyle\sum_{i=1}^{N} \left[ \textbf{p}_i(\tau) \vartheta_i(\tau) - \textbf{p}_i(0) \vartheta_i(0) \right] = 
 \nonumber \\
 - \int\limits_{0}^{\tau} \left[ \displaystyle\sum_{i = 1}^{N}  \frac{\textbf{p}_i \textbf{p}_i }{m_i} \cdot  d\textbf{S}_i  
- \frac{1}{2} \displaystyle\sum_{i,j}^{N} \textbf{f}_{ij} \vartheta_{ij}     - \displaystyle\sum_{i=1}^{N} \textbf{f}_{i_{\textnormal{ext}}}  \vartheta_i \right] dt, \nonumber
\end{align}
and using identity 
(\ref{surfaceflux}),
\begin{align}
 \overbrace{\displaystyle\sum_{i=1}^{N} \left[ \textbf{p}_i(\tau) \vartheta_i(\tau) \! - \! \textbf{p}_i(0)\vartheta_i(0) \right]}^{\text{Accumulation}} + \! \overbrace{\displaystyle\sum_{i = 1 }^{N} \displaystyle\sum_{k=1}^{N_t} \textbf{p}_i\! \sum_{\alpha=1}^3 \frac{p_{\alpha i}}{|p_{\alpha i}|} dS_{\alpha i ,k}}^{\text{Advection}}
\nonumber \\
=\underbrace{\displaystyle\sum_{i,j}^{N} \int\limits_{0}^{\tau}  \textbf{f}_{ij}(t) \vartheta_{ij}(t) dt + \displaystyle\sum_{i=1}^{N} \int\limits_{0}^{\tau} \textbf{f}_{i_{\textnormal{ext}}}(t)  \vartheta_i(t) dt}_{\text{Forcing}}.
\label{momentumfluxMD}
\end{align}
The integral of the forcing term can be 
rewritten as the sum,
\begin{align}
\int\limits_{0}^{\tau}  \textbf{f}_{ij}(t) \vartheta_{ij}(t) dt \approx  
\Delta t  \displaystyle\sum_{n=1}^{N_{\tau}} \textbf{f}_{ij} \left(t_n\right) \vartheta_{ij}\left(t_n\right),  \nonumber
\end{align}
where $N_{\tau}$ is the number time steps. Equation (\ref{momentumfluxMD}) can be rearranged as follows,
\begin{align}
\displaystyle\sum_{i=1}^{N}  \frac{ p_{\alpha i}(\tau) \vartheta_i(\tau) - p_{\alpha i}(0)\vartheta_i(0)}{\tau \Delta V} 
\nonumber \\
+ \frac{ \{\overline{\rho u_\alpha u_\beta} \}^+ - \{\overline{\rho  u_\alpha u_\beta}\}^-}{\Delta r_\beta}
= - \frac{\overline{K}_{\alpha \beta}^{\,+} - \overline{K}_{\alpha \beta}^{\,-}}{\Delta r_\beta} 
\nonumber \\
+ \frac{\overline{T}_{\alpha \beta}^{\,+} - \overline{T}_{\alpha \beta}^{\,-}}{\Delta r_\beta} + \frac{1}{N_\tau \Delta V} \displaystyle\sum_{i = 1}^{N} \displaystyle\sum_{n = 1}^{N_\tau}  f_{\alpha i_{\textnormal{ext}}}(t_n) \vartheta_i(t_n), 
\label{final_stress_equation}
\end{align}
where the overbar denotes the time average. The time-averaged traction in (\ref{final_stress_equation}) is given by,
\begin{align}
\overline{T}_{\alpha \beta}^{\,\pm} = - \frac{1}{N_\tau}  \frac{1}{4 \Delta A_\beta} \displaystyle\sum_{i,j}^{N} \displaystyle\sum_{n=1}^{N_\tau}  \textbf{f}_{\alpha ij}(t_n) dS_{\beta ij}^\pm(t_n), \nonumber 
\end{align}
The time-averaged kinetic surface pressure in (\ref{final_stress_equation}) is, 
\begin{align}
\overline{K}_{\alpha \beta}^{\,\pm}  =  \frac{1}{\tau} \frac{1}{2 \Delta A_\beta}  \displaystyle\sum_{i = 1 }^{N} \displaystyle\sum_{k=1}^{N_t}  \frac{p_{\alpha i}(t_k) p_{\beta  i}(t_k)}{|p_{\beta i}(t_k)|} d S_{\beta i ,k}^\pm(t_k)
\nonumber \\
 - \{\overline{\rho u_\alpha u_\beta} \}^\pm. \nonumber 
\end{align}
The \eq{final_stress_equation} demonstrates that the time average of 
the fluxes, stresses and body forces on a CV during the interval $0$ to $\tau$, 
completely determines the change in momentum within the CV for a single trajectory of the system through phase space (i.e. an MD simulation). 
The time evolution of the microscopic system, Eq. (\ref{final_stress_equation}), can also be obtained directly by
evaluating the derivatives of the mesoscopic expression (\ref{mesoevo_traction}) and invoking the ergodic hypothesis, hence
replacing  $\big\langle\alpha ;\mathit{f\big\rangle}$ with$~\frac{1}{\tau}\int_{0}^{\tau }\alpha dt$. The use of the ergodic
hypothesis is justified provided that the time interval, $\tau $, is
sufficient to ensure phase space is adequately sampled.

Finally, there are no new techniques required to integrate the energy Eq. \ref{microfluxE},
 \begin{align}
\displaystyle\sum_{i=1}^{N} \left[ e_i(\tau) \vartheta_i(\tau) - e_i(0) \vartheta_i(0) \right] 
\nonumber \\
 =  - \int\limits_0^\tau \left[ \! \displaystyle\sum_{i=1}^{N} \! e_i \frac{\textbf{p}_i}{m_i}  \cdot  d\textbf{S}_i
 - \frac{1}{2}\displaystyle\sum_{i,j}^{N} \frac{\textbf{p}_i}{m_i}  \cdot \textbf{f}_{ij}  \vartheta_{ij} \right] dt
\end{align}
which gives the final form, written without external forcing,
 \begin{align}
\overbrace{\displaystyle\sum_{i=1}^{N} \left[ e_i(\tau) \vartheta_i(\tau) \! - \! e_i(0) \vartheta_i(0) \right] }^{\text{Accumulation}}
\! + \! \overbrace{\displaystyle\sum_{i = 1 }^{N} \displaystyle\sum_{k=1}^{N_t} e_i  \! \sum_{\alpha=1}^3 \frac{ p_{\alpha i}}{|p_{\alpha i}|} dS_{\alpha i ,k}}^{\text{Advection}}
 \nonumber \\
= \underbrace{ \frac{1}{2}\displaystyle\sum_{i,j}^{N}  \int\limits_0^\tau \frac{\textbf{p}_i(t)}{m_i}  \cdot \textbf{f}_{ij}(t)  \vartheta_{ij}(t) dt }_{\text{Forcing}}. 
\label{energyfluxMD}
\end{align}
As in the momentum balance equation, the integral of the forcing term can be 
approximated by the sum,
\begin{align}
\int\limits_{0}^{\tau} \frac{\textbf{p}_i(t)}{m_i}  \cdot  \textbf{f}_{ij}(t) \vartheta_{ij}(t) dt 
\nonumber \\
\approx  \Delta t  \displaystyle\sum_{n=1}^{N_{\tau}} \frac{\textbf{p}_i(t_n)}{m_i}  \cdot \textbf{f}_{ij} \left(t_n\right) \vartheta_{ij}\left(t_n\right),  \nonumber
\end{align}
where $N_{\tau}$ is the number time steps. 

In the next section, the elements, Accumulation, Advection and Forcing in the above equations
are computed individually in an MD simulation to confirm Eqs. (\ref{massfluxMD}), (\ref{momentumfluxMD}) and (\ref{energyfluxMD}) numerically.

\abovesubsectionspace
\subsection{Results and Discussion}
\belowsubsectionspace
\label{sec:MD_results_discussion}

Molecular Dynamics (MD) simulations in 3D are used in this section to validate numerically, and explore the 
statistical convergence of, the CV formalism for three test cases. 
The first investigation was to confirm numerically the conservation properties of an arbitrary control volume. 
The second simulation compares the value of the scalar pressure obtained from the molecular CV formulation with 
that of the virial expression for 
an equilibrium system in a periodic domain.
The final test is a Non Equilibrium Molecular Dynamics (NEMD) simulation of the start-up of Couette flow 
initiated by translating the top wall in a slit channel geometry.
The NEMD system is analyzed using the CV expressions Eqs.\ (\ref{massfluxMD}), (\ref{momentumfluxMD}) and (\ref{energyfluxMD}),
and the shear pressure was computed by the VA and CV routes.
Newton's equations of motion were integrated using the half-step leap-frog Verlet algorithm, \cite{Allen_Tildesley}.
The repulsive Lennard-Jones (LJ) or Weeks-Chandler-Anderson (WCA) potential \citep{Rapaport},
\begin{align}
        \varPhi(r_{ij})=4\epsilon \left[ \left(  \frac{\ell}{r_{ij}} \right) ^{12} - \left(  \frac{\ell}{r_{ij}} \right) ^{6} \right] + \epsilon , \; \;  r_{ij}\leq r_c,
\label{LJpot}
\end{align}
was used for the molecular interactions, which is the Lennard-Jones potential shifted upwards 
by $\epsilon$ and truncated at the minimum in the potential, $r_{ij} = r_c \equiv 2^{1/6} \ell$. 
The potential is zero for $r_{ij} > r_{c}$. The energy scale is set by $\epsilon$, the length scale by $\ell$ and molecular mass by $m$.
The results reported here are given in terms of $\ell, \epsilon$ and $m$. A timestep of $0.005$ was used for all simulations. The domain size in the first two simulations was $13.68$, which contained $N=2048$ molecules, the density was $\rho = 0.8$ and the reduced temperature was set to an initial value of $T = 1.0$. Test cases 1 and 2 described below are for equilibrium systems, and therefore did not require thermostatting. Case 3 is for a non-equilibrium system and required removal of generated heat, which was achieved by thermostatting the wall atoms only.
\vspace{-0.1in}
\subsubsection{Case 1}
\begin{figure}
  \subfigure{\label{fig:CV_evolutionM}\includegraphics[width=0.47\textwidth]{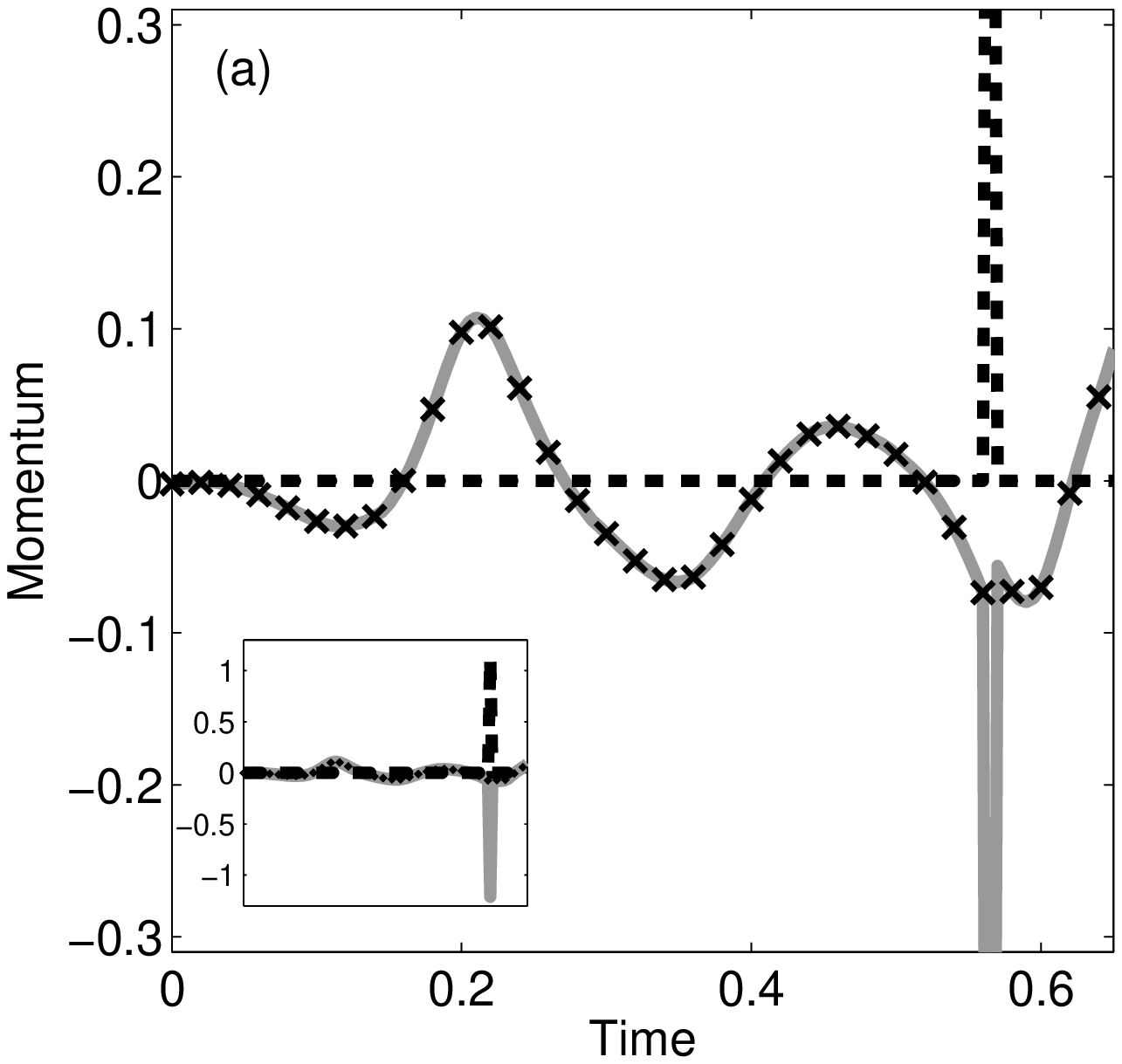}} \\
  \subfigure{\label{fig:CV_evolutionE}\includegraphics[width=0.47\textwidth]{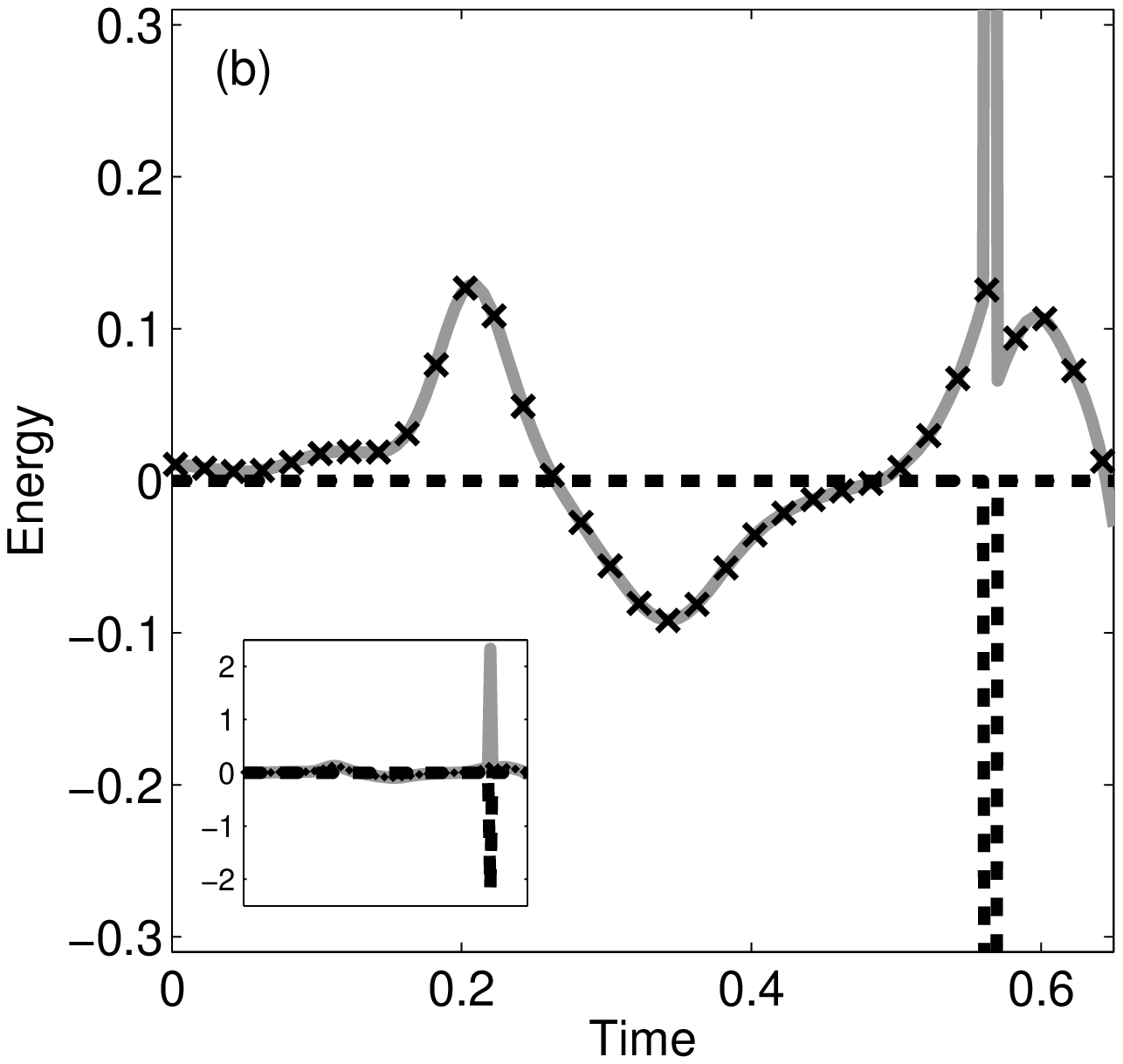}}
 \caption{The various components in Eq.\ \ref{residual}, `Accumulation' (\textcolor{gray}{---}), the 
time integral of the surface force, `Forcing' ($\boldsymbol{\times}$), and momentum flux term, `Advection' (- - -) are shown. `Forcing' symbols are shown every 4th timestep for clarity and the insert shows the full ordinate scale over the same time interval on the abscissa. From top to bottom, (a) Momentum Control Volume, (b) Energy Control Volume.}
\figspace
\end{figure}

\vspace{-0.15in}
In case 1, the periodic domain simulates a constant energy ensemble. The separate terms of the integrated mass, momentum and energy equations given in (\ref{massfluxMD}), (\ref{momentumfluxMD}) and (\ref{energyfluxMD}) were evaluated numerically for several sizes of CV. The mass conservation can readily be shown to be satisfied as it simply requires tracking the number of molecules in the CV. The momentum and energy balance equations are conveniently checked for compliance at all times by evaluating the residual quantity,
\begin{align}
Residual = \text{Accumulation} - \text{Forcing} + \text{Advection},
\label{residual}
\end{align}
which must be equal to zero at all times for the CV equations to be satisfied. 
This was demonstrated to be the case, as may be seen in Figs.\ \ref{fig:CV_evolutionM} and \ref{fig:CV_evolutionE}, for 
a cubic CV of side length $1.52$ in the absence of body forces. 
The evolution of momentum inside the CV is shown numerically to be exactly equal to 
the integral of the surface forces until a molecule 
crosses the CV boundary. Such events give rise to a momentum flux contribution which appears as a spike in the 
Advection and Accumulation terms, as is evident in Fig.\ \ref{fig:CV_evolutionM}. The residual nonetheless 
remains identically zero (to machine precision) at all times.
The energy conservation is also displayed in Fig.\ \ref{fig:CV_evolutionE}. The average error over the period of the simulation ($100$ MD timeunits) was less than 1\%, where the average error is defined as the ratio of the mean $|Residual|$ to the mean $|Accumulation|$ over the simulation. The error is attributed to the use of the leapfrog integration scheme, a conclusion supported by the linear decrease in error as timestep $\Delta t \rightarrow 0$.
\vspace{-0.3in}
\subsubsection{Case 2}
\begin{figure}
\includegraphics[width=0.48\textwidth]{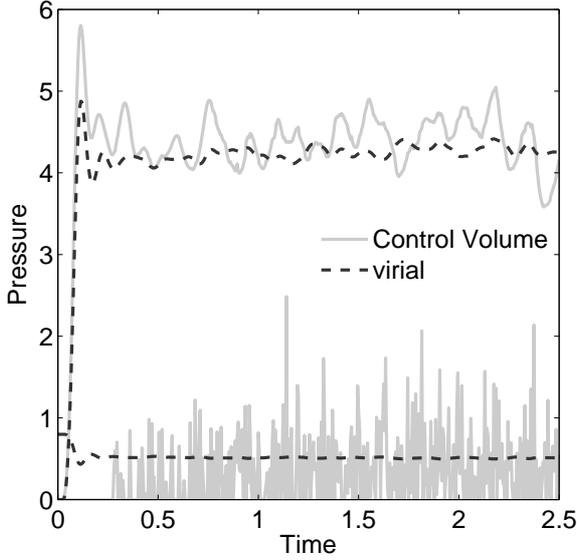}
\caption{\label{fig:Virial_vs_CV} 
$\Pi_{vir}$ and $\Pi_{CV}$ from  Eqs.\ (\ref{pressure_scalar_virial}) 
and (\ref{pressure_scalar_CV}) respectively.  The configurational and kinetic pressures are separated with configurational values typically having greater magnitudes ($\sim4.0$) than kinetic ($\sim0.6$).
Continuous lines are control volume pressures and dotted lines are virial pressure.}
\figspace
\end{figure}

\begin{figure}
\includegraphics[width=0.48\textwidth]{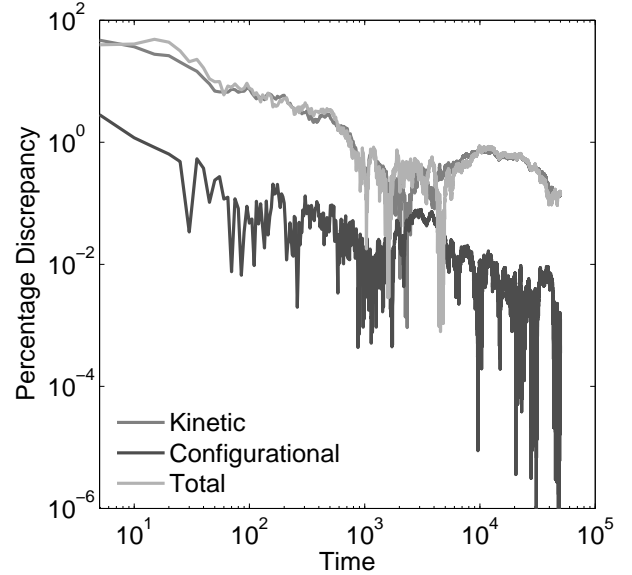}
\caption{\label{fig:Virial_vs_CV_error} 
The percentage relative difference between the virial and control volume time-accumulated 
scalar pressures (PD defined in the text). 
Values for the kinetic, configurational and total PD are shown.}
\figspace
\end{figure}

\vspace{-0.15in}
As in case 1, the same periodic domain is used in case 2 to simulate a constant energy ensemble.
The objective of this exercise is to show that the average of the virial formula for the 
scalar pressure, $\Pi_{vir}$, applicable to an equilibrium periodic system, 
\begin{align}
\Pi_{vir} = \frac{1}{3V}\displaystyle\sum_{i=1}^{N} 
\bigg\langle \frac{ \overline{\textbf{p}}_i \cdot \overline{\textbf{p}}_{i}}{m_i}  
+ \frac{1}{2} \displaystyle\sum_{i\neq j}^{N} \textbf{f}_{ij} \cdot \textbf{r}_{ij} ; \textit{f} \bigg\rangle,
\label{pressure_scalar_virial}
\end{align}
arises from the intermolecular interactions across the periodic boundaries \citep{Tsai}. The CV formula for the scalar pressure is,
\begin{align}
\!\!\!\! \Pi_{CV} = \!\! \frac{1}{6} \! \left(P_{xx}^{+} \! + \! P_{xx}^{-} \! + \! P_{yy}^{+} \! + \! P_{yy}^{-} \! + \! P_{zz}^{+} \! + \! P_{zz}^{-} \right),
\label{pressure_scalar_CV}
\end{align}
where the $P_{\alpha\alpha }^{\pm}$ normal pressure is defined in \eq{CV_pressure} and includes 
both the kinetic and configurational components on each surface. Both routes involve the pair forces, $f_{ij}$. However, the CV expression which uses MOP counts only those pair forces which cross a plane while VA (Virial) 
sums $f_{ij} r_{ij}$ over the whole volume. It is therefore expected that there would be differences between the two methods at short times, converging at long times. A control volume the same size as the periodic box was taken.
The time averaged control volume, ($\Pi_{CV}$) and virial ($\Pi_{vir}$) pressure values are shown in 
Fig.\ \ref{fig:Virial_vs_CV} to converge towards the same value with increasing 
time. The simulation is started from an FCC lattice with a short range potential (WCA) so the initial configurational stress is zero. It is the evolution of the pressure from this initial state that is compared in Fig.\ \ref{fig:Virial_vs_CV}.
The virial kinetic pressure makes use of the instantaneous values of the domain molecule's 
velocities at every time step. In contrast, the CV kinetic part of the pressure is due to molecular surface crossings only, which may explain its slower convergence to the limiting value than the kinetic part of the virial expression. To quantify this difference in convergence for the two measures of the pressure, the standard deviation, $\text{SD}(x)$, is evaluated, ensuring decorrelation \citep{Delgado-Buscalioni_DeFabritiis} using block averaging \citep{Rapaport}. For the kinetic virial, $\text{SD}(\kappa_{vir})= 0.0056$, and configurational, $\text{SD}(\sigma_{vir}) = 0.0619$. For the kinetic CV pressure $\text{SD}(\kappa_{CV})= 0.4549$ and configurational $\text{SD}(\sigma_{CV}) = 0.2901$. The CV pressure, which makes use of the MOP formula, would therefore require more samples to converge to a steady state value. However, the MOP pressures are generally more efficient to calculate than the VA. More usefully, from an evaluation of only the interactions over the outer CV surface, the pressure in a volume of arbitrary size can be determined.  \\
%
Figure\ \ref{fig:Virial_vs_CV_error} is a log-log plot of the Percentage
Discrepancy (PD) between the two ($PD=\left[100 \times |\Pi_{CV}-\Pi_{vir}|/\Pi_{vir}\right]$). 
After $10$ million timesteps or a reduced time of $5 \times 10^{4}$, the percentage discrepancy 
in the configurational part has decreased to $0.01\%$, and the kinetic part of the 
pressure matches the virial (and kinetic theory) to within $0.1\%$. The total pressure value 
agrees to within $0.1\%$ at the end of this averaging period.
The simulation average temperature was $0.65$, and
the kinetic part of the CV pressure was statistically the same as 
the kinetic theory formula prediction, $\kappa_{CV} =\rho k_B T=0.52$ \citep{Rapaport}.
The VA formula for the pressure in a volume the size of the domain is by definition formally
the same as that of the  virial pressure. 
The next test case compares the CV and VA formulas for the shear stress in a system out of equilibrium. \\

\vspace{-0.3in}
\subsubsection{Case 3}
\begin{figure}
\includegraphics[width=0.463\textwidth]{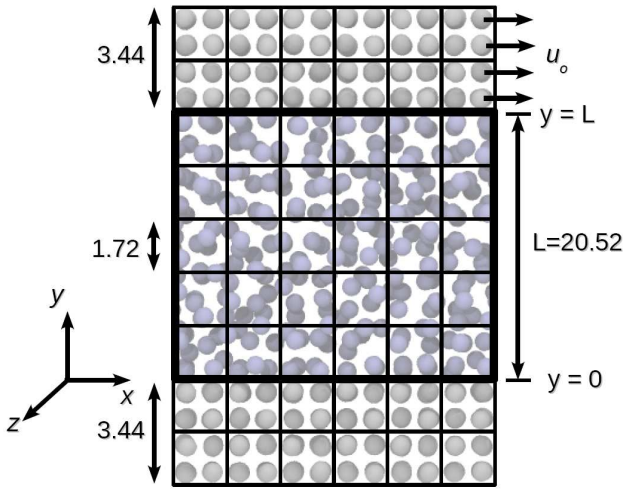}
\caption{\label{fig:couette_MD_CV} (Color online) Schematic diagram of the  NEMD simulation geometry consisting of a 
sliding top wall and stationary bottom wall, both composed of tethered atoms. 
The simulation domain contained a lattice of contiguous CV used for pressure averaging 
(shown by the small boxes) while the thicker 
line denotes a single CV containing the entire liquid region.}
\figspace
\end{figure}

\vspace{-0.1in}
In this simulation study,  
Couette flow was simulated by entraining a model liquid between two solid walls. 
The top wall was set in translational motion parallel to the bottom (stationary) wall and 
the evolution of the velocity profile towards the  steady-state Couette flow limit was followed.
The velocity profile, and the derived CV and VA shear stresses are compared with the analytical solution 
of the unsteady diffusion equation.
Four layers of 
tethered molecules were used to represent each wall, 
with the top wall given a sliding velocity of, $U_0 = 1.0$ at the start of the simulation, time $t=0$. 
The temperature of both walls was controlled by applying the Nos\'{e}-Hoover (NH) thermostat 
to the wall atoms \citep{Hoover}. The two walls were thermostatted separately, and 
the equations of motion of the wall atoms were,
\begin{subequations}
 \begin{align}
\dot{\textbf{r}}_{i} = \frac{\overline{\textbf{p}}_{i}}{m_i} + U_0 \textbf{n}_x^+, \\
\dot{\overline{\textbf{p}}}_{i} = \textbf{F}_{i} + \textbf{f}_{i_{\textnormal{ext}}} - \xi \overline{\textbf{p}}_{i}, \\
\textbf{f}_{i_{\textnormal{ext}}} = \textbf{r}_{i_0}  \left( 4 k_{4} r_{i_0}^{2}+6 k_{6} r_{i_0}^{4} \right), \\
\dot{\xi} = \frac{1}{Q_\xi} \left[  \displaystyle\sum_{n=1}^{N}  \frac{\overline{\textbf{p}}_{n} \cdot \overline{\textbf{p}}_{n}}{m_n} - 3T_0 \right], \label{NH}
 \end{align}
\end{subequations}
where $\textbf{n}_x^+$ is a unit vector in the $x-$direction, $m_{n}\equiv m$, and
$\textbf{f}_{i_{\textnormal{ext}}}$ is the tethered atom force, using 
the formula of \citet{Petravic_Harrowell} 
($k_{4}=5 \times 10^{3}$ and $k_{6}=5 \times 10^{6}$).
The vector, $\textbf{r}_{i_0} = \textbf{r}_{i} - \textbf{r}_0 $, is the displacement of 
the tethered atom, $i$, from its lattice site coordinate, $\textbf{r}_0$.
The Nos\'{e}-Hoover thermostat dynamical variable is denoted by $\xi$, $T_0 = 1.0$ 
is the target temperature of the wall, and
the effective time constant or damping coefficient,  in \eq{NH} was given the 
value, $Q_\xi = N \Delta t$.
The simulation was carried out for a cubic domain of sidelength $27.40$, of which 
the fluid region extent was $20.52$ in the $y-$direction. Periodic boundaries were used 
in the streamwise ($x$) and spanwise ($z$) directions. The results presented 
are the average of eight simulation trajectories starting 
with a different set 
of initial atom velocities. The lattice contained $16384$ molecules and was at a density of $\rho = 0.8$.
The molecular simulation domain was sub-divided into $4096$ ($16^3$) control volumes, and  the average velocity and 
shear stress was determined in each of them. A larger single CV encompassing all of
the liquid region of the domain, shown bounded by the thick line  in Fig. \ref{fig:couette_MD_CV},
was also considered.

\begin{figure}
\includegraphics[width=0.48\textwidth]{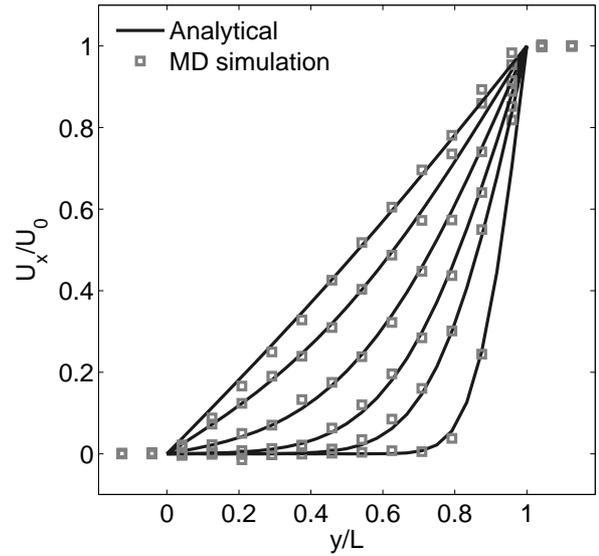}
\caption{\label{fig:velocity_slice} The $y-$ dependence of the streaming velocity profile at  
times $t=2^n$ for $n=0,2,3,4,5,6$ from right to left. The squares are the NEMD CV data values
and the analytical solution to the continuum equations of \eq{analytical_couette} is
given at the same  six times as continuous curves.}
\figspace
\end{figure}

The continuum solution for this configuration is considered now.
Between two plates, 
there are no body forces and the flow eventually becomes fully developed, \citep{Potter_Wiggert} so that 
Eq.\ (\ref{BofMEqn2}) can be simplified and after applying the divergence 
theorem from Eq.\ (\ref{divergence}) it becomes,
 \begin{align}
 \frac{\partial }{\partial t}  \int_V  \rho \boldsymbol{u} dV = - \int_V \boldsymbol{\nabla} \cdot \boldsymbol{\Pi} dV,   \nonumber
 \end{align}
which is valid for any arbitrary volume in the domain 
and must be valid at any point for a continuum. 
The shear pressure in the fluid, $\Pi_{xy}(y)$, drives the time evolution,
 \begin{align}
 \frac{\partial \rho u_x }{\partial t}   = - \frac{\partial \Pi_{xy}}{\partial y}. \nonumber
 \end{align}
For a Newtonian liquid with viscosity, $\mu$, \citep{Potter_Wiggert},
\begin{align}
\Pi_{xy} = - \mu \frac{\partial u_x}{\partial y},
\label{continuum_stress}
\end{align}
this gives  the 1D diffusion equation,
\begin{align}
\frac{\partial u_x}{\partial t} = \frac{\mu}{\rho} \frac{\partial^2 u_x}{\partial y^2},
\label{diffusion}
\end{align} 
assuming the liquid to be incompressible.
This can be solved for the boundary conditions, 
\begin{align}
u_x(0,t) = 0 \;\;\;\;\;\; u_x(L,t) = U_0 \;\;\;\;\;\; u_x(y,0) = 0, \nonumber
\end{align}
where the bottom and top wall-liquid boundaries are at $y = 0$ and $y = L$, respectively. 
The Fourier series solution of these equations with inhomogeneous boundary conditions \citep{Strauss} is,
\begin{align}
  u_x(y,t)=%
  \begin{cases}
    \;\;\;\;\;\;\;\;\;\;\; U_0 & \quad y=L\\
     \displaystyle\sum_{n=1}^{\infty} u_n(t)  sin \left(\frac{n \pi y}{L}\right)   & \quad 0<y<L\\
    \;\;\;\;\;\;\;\;\;\;\;\; 0   & \quad y=0\\ 
  \end{cases}
\label{analytical_couette}
\end{align}
where  
$\lambda_n = (n \pi/L)^{2}$
and $u_n(t)$ is given by,
\begin{align}
u_n(t) = \frac{2 U_0 (-1)^n  }{n \pi}\left[ \exp \left({- \frac{\lambda_n\mu t}{\rho} } \right) -1 \right]. \nonumber
\end{align}
The velocity profile resolved at the control volume level is compared with 
the continuum solution in Fig.\ \ref{fig:velocity_slice}. There were 
$16$ cubic NEMD CV of side length $1.72$ spanning the system in the $y$ direction, 
with each data point on the figure being derived from 
a local time average of $0.5$ time units. The analytic continuum solution was 
evaluated numerically from Eq.\ (\ref{analytical_couette}) with $n=1000$ and $\mu = 1.6$, 
the latter a literature value for the WCA fluid shear viscosity at $\rho =0.8$ and $T=1.0$, 
\citep{Da_Costa_Silva_et_al}.
There is mostly very good agreement between the analytic and NEMD velocity profiles at all times, 
although some effect of the stacking of molecules near the two walls can be seen in a slight blunting 
of the fluid velocity profile very close to the tethered walls (located by the horizontal two squares on the 
far left and right of the figure) which is an aspect of the molecular system that 
the continuum treatment is not capable of reproducing.\\

\begin{figure}
\includegraphics[width=0.48\textwidth]{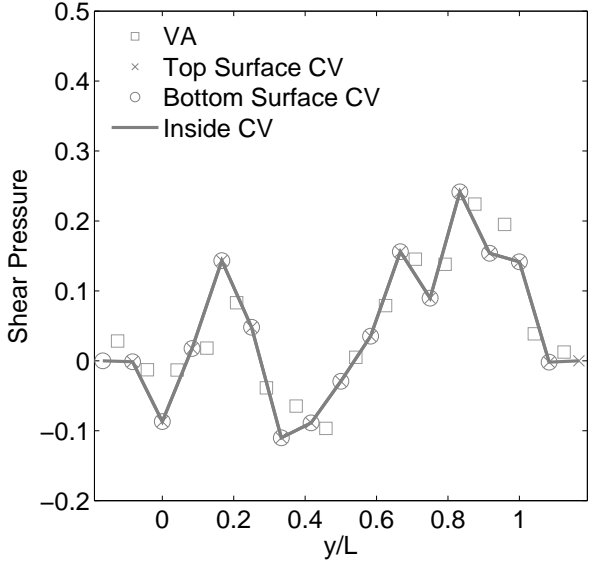}
\caption{\label{fig:VA_vs_CV_slice} 
The $y-$dependence of the shear pressure at $t=10$, averaged over $100$ timesteps
and for a single simulation trajectory.  
The VA value from \eq{VA_pressure} are the squares. The CV surface traction
from \eq{CV_pressure} is indicated by $\times$ and $\circ$ for 
the top and bottom surfaces, respectively. The solid gray line displays the resulting pressure field using \eq{FEA} with linear shape functions.} 
\figspace
\end{figure}

The VA and CV shear pressure, given by Eqs.\ (\ref{VA_pressure}) and (\ref{CV_pressure}), 
are compared at time $t=10$ in Fig. \ref{fig:VA_vs_CV_slice}. 
The comparison is for a single simulation trajectory  
resolved  into $16$ cubic volumes of size $1.72$ in the $y-$direction, with averaging in 
the $x$ and $z$ directions and over $0.5$ in reduced time.
The figure shows the shear pressure on the faces of the CV. Inside the CV, the pressure 
was assumed to vary linearly, and the value at the midpoint is shown to be 
comparable to the VA-determined value.
Figure \ref{fig:VA_vs_CV_slice} shows that there is  good agreement between the VA and CV approaches. 
Note that the CV pressure is effectively the MOP formula applied to the 
faces of the cube, and hence this case study demonstrates a consistency between MOP and VA.
We have shown previously that this is true for the special case of an infinitely thin bin or 
the limit of the pressure at a plane \citep{Heyes_et_al}. Practically, the extent of agreement 
in this exercise is limited 
by the inherent assumptions and spatial resolution of the two methods;
a single average over a volume is required for  VA, but a 
linear pressure relationship is assumed for  CV to obtain the pressure 
tensor value corresponding to the center of the CV. \\

The continuum analytical $xy$ pressure tensor component 
can be derived analytically using the same Fourier series approach 
for $\partial u_x/\partial y$,\citep{Strauss},
\begin{align}
\Pi_{xy}(y,t) = - \frac{\mu U_0}{ L } \!\! \left[1 + 2  \!\! \displaystyle\sum_{n=1}^{\infty} (-1)^n e^{- \frac{\lambda_n \mu t}{\rho} }  cos \left( \frac{n \pi y}{L} \right) \right], 
\label{analytical_stress}
\end{align}
which is valid for the entire domain $0 \leq y \leq L$. 

\begin{figure}
\includegraphics[width=0.48\textwidth]{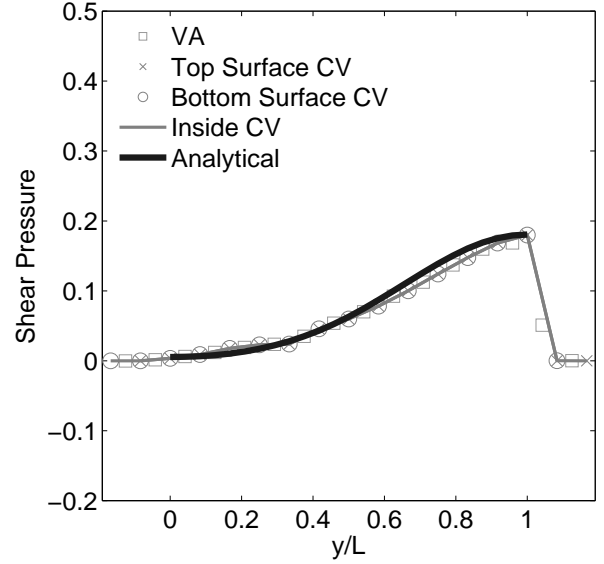}
\caption{\label{fig:VA_vs_CV_slice_SS} 
As Fig.\ \ref{fig:VA_vs_CV_slice}, except that the NEMD results are averaged over a
set of eight independent simulations of $1,000$ timesteps ($5$ reduced 
time units) each. The simulation-derived VA and CV shear pressures are compared with the continuum analytical 
solution given in \eq{analytical_stress} (solid black line). The jump in the profile 
on the right of the figure is due to the presence of the tethered wall.}
\figspace
\end{figure}

A statistically meaningful comparison between the CV, VA and 
continuum analytic shear pressure profiles requires more 
averaging of the simulation data than for the streaming velocity, \citep{Hadjiconstantinou_et_al}, and eight 
independent simulation trajectories over $5$ reduced time units were used. Figure\ \ref{fig:VA_vs_CV_slice_SS} shows that the three methods exhibit 
good agreement within the simulation statistical uncertainty.\\

\begin{figure}
\includegraphics[width=0.48\textwidth]{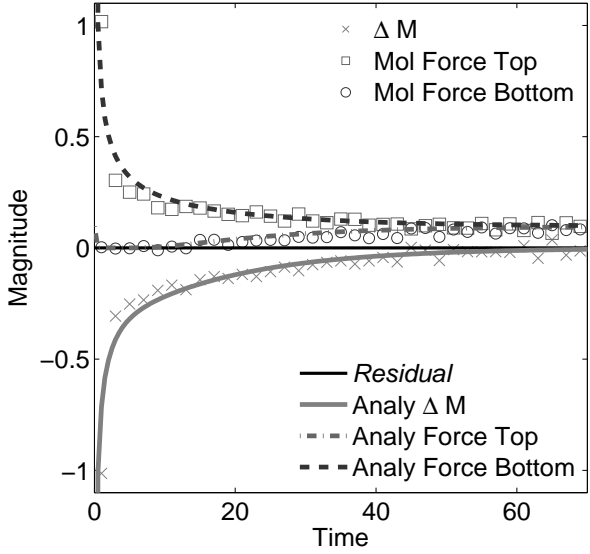}
\caption{\label{fig:couette_CV} The evolution of surface forces and 
momentum change for a molecular CV from Eq.\ (\ref{CV_couette}), 
(points) and analytical solution for the 
continuum (Eqs.\ (\ref{analytical_CVftop}), (\ref{analytical_CVfbottom}) and (\ref{analytical_CVmomevo})), 
presented as lines on the figure. The $Residual$, defined in \eq{residual}, is also given.
Each point represents the average over an ensemble of eight independent systems and 
40 timesteps.}
\figspace
\end{figure}

As a final demonstration of the use of the  CV equations, the control volume is now 
chosen to encompass the entire liquid domain (see Fig.\ \ref{fig:couette_MD_CV}), 
and therefore the external forces arise from interactions with the wall atoms only.
The momentum equation,  Eq.\ (\ref{microfluxM}), is written as,
 \begin{align}
  \frac{\partial}{\partial t}\displaystyle\sum_{i=1}^{N}  \textbf{p}_i \vartheta_i =  -\displaystyle\sum_{i = 1}^{N}   \overbrace{ \frac{\textbf{p}_i \textbf{p}_i }{m_i} \cdot  d\textbf{S}_i }^{\circlenum{1}}  + \displaystyle\sum_{i=1}^{N} \overbrace{\textbf{f}_{i_{\textnormal{ext}}}  \vartheta_i}^{\circlenum{3}}.
 \nonumber \\  
 - \frac{1}{2} \displaystyle\sum_{i,j}^{N}  \big[ \underbrace{\textbf{f}_{ij} dS_{xij}}_{\circlenum{2}} + \underbrace{\textbf{f}_{ij}}_{\circlenum{4}} dS_{yij} + \underbrace{\textbf{f}_{ij} dS_{zij}}_{\circlenum{2}}\big], \nonumber
 \end{align}
which can be simplified as follows. For term, $\circlenum{1}$ in the above equation, the 
fluxes across the CV boundaries in the streamwise and spanwise directions cancel due to the 
periodic boundary conditions. Fluxes across the $xz$ boundary surface are zero as the 
tethered wall atoms prevent such crossings. The force term, $\circlenum{2}$, also 
vanishes because across the periodic boundary, 
$\textbf{f}_{ij}dS_{xij}^+ =-\textbf{f}_{ij}dS_{xij}^- $,  (similarly for $z$).
The external force term, $\circlenum{3}$, is zero because all the forces in the 
system result from interatomic interactions. The sum of the  $f_{yij}$ force
components across 
the horizontal boundaries will be equal and opposite, and by symmetry the 
two $f_{zij}$ terms in $\circlenum{4}$ will be zero on average.
The above equation therefore reduces to,
\begin{align}
 \frac{\partial}{\partial t}\displaystyle\sum_{i=1}^{N}  \textbf{p}_i \vartheta_i =  
- \frac{1}{2} \displaystyle\sum_{i,j}^{N} \left[ f_{xij} dS_{yij}^+ - f_{xij} dS_{yij}^-\right].
\label{CV_couette}
\end{align}
As the simulation approaches steady state, the rate of change of momentum in the control volume 
tends to zero because the difference between the shear stresses acting across 
the top and bottom walls vanishes. The forces on the $xz$ plane boundary 
and momentum inside the CV are plotted in  Fig.\ \ref{fig:couette_CV} to
confirm \eq{CV_couette} numerically.
The time evolution of these molecular momenta and surface stresses are compared to 
the analytical continuum solution for the CV,
\begin{align}
\frac{\partial }{\partial t}  \int_V  \rho u_x dV = - \left[ \int_{S_{f}^+} \Pi_{xy} dS_{f}^+ - \!\! \int_{S_{f}^-} \Pi_{xy} dS_{f}^- \right].
\label{continuum_couette_CVstress}
\end{align}
The normal components of the pressure tensor are non-zero in the continuum, but exactly 
balance across opposite CV faces, i.e. $\Pi_{xx}^+ = \Pi_{xx}^-$. By appropriate choice of the gauge 
pressure, $\Pi_{xx}$ does not appear in the governing Eq. (\ref{continuum_couette_CVstress}).
The left hand side of the above equation is evaluated from the analytic expression for $u_{x}$,
\begin{align}
 \frac{\partial }{\partial t}  \int_V  \rho u_x dV=  2 \Delta x \Delta z \frac{\mu U_0 }{ L} \displaystyle\sum_{n=1}^{\infty}  \left[ 1 - (-1)^n \right] e^{- \frac{\lambda_n \mu t}{\rho} }.
\label{analytical_CVmomevo}
\end{align}
The right hand side is obtained from the analytic continuum expression for the
shear stress, for the bottom surface at $y=0$, 
\begin{align}
 \int_{S_{f}^+} \Pi_{xy} dS_{f}^+ = -2 \Delta x \Delta z \frac{\mu U_0}{L} \displaystyle\sum_{n=1}^{\infty}  e^{- \frac{\lambda_n \mu t}{\rho} },
\label{analytical_CVftop}
\end{align}
and for the top $y=L$,
\begin{align}
\! \int_{S_{f}^-} \Pi_{xy} dS_{f}^-= -2 \Delta x \Delta z \frac{\mu U_0}{L} \displaystyle\sum_{n=1}^{\infty} (-1)^{n} e^{- \frac{\lambda_n \mu t}{\rho} }.
\label{analytical_CVfbottom} 
\end{align}
In  Fig \ref{fig:couette_CV}, the momentum evolution on the left hand side of Eq.\ (\ref{CV_couette}) 
is compared to \eq{analytical_CVmomevo}. Equations\ (\ref{analytical_CVftop}) 
and (\ref{analytical_CVfbottom}) are also given for the shear stresses acting across the top and 
bottom of the molecular control volume (right hand side of \eq{CV_couette}).
The scatter seen in the MD data reflects the thermal fluctuations in the forces and 
molecular crossings of the CV boundaries. The average response nevertheless agrees well with the 
analytic solution, bearing in mind the element of uncertainty 
in the matching state parameter values.
This example demonstrates the potential of the CV approach applied on 
the molecular scale, as it can be seen that computation of the forces across the CV 
boundaries determines completely the average molecular microhydrodynamic response 
of the system contained in the CV. In fact, the force on only one of the surfaces 
is all that was required, as the force terms for the opposite surface could have
been obtained from \eq{CV_couette}.\\

\abovesectionspace
\abovesectionspace
\abovesectionspace
\section{Conclusions}
\belowsectionspace

In analogy to continuum fluid mechanics, the evolution equations for a
molecular systems has been expressed within a Control Volume (CV) in terms of 
fluxes and stresses across the surfaces.  
A key ingredient is the definition and manipulation of a Lagrangian to Control Volume 
conversion function, $\vartheta$, which identifies molecules within the CV.
The final appearance of the equations has the same form as Reynolds' 
Transport Theorem applied to a discrete system. The equations presented follow directly from Newton's equation of motion for a system of discrete particles, requiring no additional assumptions and therefore sharing the same range of validity.

Using the \CV~function, the relationship between Volume Average (VA) \citep{Lutsko,Cormier_et_al} and Method Of Planes (MOP) pressure \citep{Todd_et_al_95, Han_Lee} has been established, without Fourier transformation. The two definitions of pressure are shown numerically to give equivalent results away from equilibrium and, for homogeneous systems, shown to equal the virial pressure.

A Navier--Stokes-like equation was derived for the evolution of momentum within 
the control volume, expressed in terms of surface fluxes and stresses.  
This provides an exact mathematical relationship between molecular fluxes/pressures and the evolution of momentum and energy in a CV. 
Numerical evaluations of the terms in the conservation of mass, momentum and energy equations demonstrated consistency with theoretical predictions.

The CV formulation is general, and can be applied to derive conservation equations for any fluid dynamical property localised to a region in space.  It can also facilitate the derivation of conservative numerical schemes for MD, and the evaluation of the accuracy of numerical schemes.  Finally, it allows for accurate evaluation of macroscopic flow properties, in a manner consistent with the continuum conservation laws.

\appendix
\abovesectionspace
\abovesectionspace
\abovesectionspace
\section{Discrete form of Reynolds' Transport Theorem and the Divergence Theorem}
\belowsectionspace
\label{sec:divergence_ReynoldsTT}

In this appendix, both Reynolds' Transport Theorem and the Divergence Theorem for a discrete system are derived. The relationship between an advecting and fixed control volume is shown using the concept of peculiar momentum. 

The microscopic form of the continuous Reynolds' Transport Theorem \citep{Reynolds} is derived for a property $\boldsymbol{\chi}=\boldsymbol{\chi}(\textbf{r}_i,\textbf{p}_i,t)$ which could be mass, momentum or the pressure tensor. The \CV \ function, $\vartheta_i$, is dependent on the molecule's coordinate; the location of the cube center, $\textbf{r}$, and side 
length, $\Delta \textbf{r}$, which are all a function of time. The time evolution of the CV is therefore,
\begin{align}
 \frac{d}{d t}\displaystyle\sum_{i=1}^{N}  \boldsymbol{\chi}(t) \vartheta_i(\textbf{r}_i(t),\textbf{r}(t), \Delta \textbf{r}(t)) 
 \nonumber \\
 =  \displaystyle\sum_{i = 1}^{N} \left[   \frac{d\boldsymbol{\chi} }{d t}   \vartheta_i + \boldsymbol{\chi} \frac{d  \textbf{r}_i}{d t} \cdot  \frac{\partial  \vartheta_i}{\partial \textbf{r}_i}  \;\;\;\;\;\;
\nonumber \nl 
+ \boldsymbol{\chi} \frac{d  \textbf{r}}{d t} \cdot  \frac{\partial  \vartheta_i}{\partial \textbf{r}}  + \boldsymbol{\chi} \frac{d  \Delta \textbf{r}}{d t} \cdot  \frac{\partial  \vartheta_i}{\partial \Delta \textbf{r}}  \right]. \nonumber
\end{align}
The velocity of the moving volume is defined as $\tilde{\boldsymbol{u}} = d  \textbf{r}/d t $, which can be different to the macroscopic velocity $\boldsymbol{u}$. Surface translation or deformation of the cube, $\partial \vartheta_i/\partial \Delta \textbf{r}$, can be included in the expression for velocity $\tilde{\boldsymbol{u}}$. The above analysis is for a microscopic system, although a similar process for a mesoscopic system can be applied and includes terms for CV movement in Eq. (\ref{EqIK18}). 

Hence Reynolds treatment of a continuous medium \citep{Reynolds} is extended here to a discrete molecular system,
\begin{align}
 \frac{d}{d t}\displaystyle\sum_{i=1}^{N}  \boldsymbol{\chi}(t) \vartheta_i(\textbf{r}_i(t),\textbf{r}(t), \Delta \textbf{r}(t)) 
\nonumber  \\ 
 =  \displaystyle\sum_{i = 1}^{N} \left[   \frac{d \boldsymbol{\chi} }{d t} \vartheta_i + \boldsymbol{\chi} \left(\tilde{\boldsymbol{u}} - \frac{\textbf{p}_i}{m_i} \right) \cdot d\textbf{S}_i \right]. 
\label{RTT} 
\end{align}
The conservation equation for the mass,  $\boldsymbol{\chi} = m_i$, in a moving reference frame is,
\begin{align}
 \frac{d}{d t}\displaystyle\sum_{i=1}^{N}  m_i \vartheta_i
 = \displaystyle\sum_{i = 1}^{N} \left[ m_i \left( \tilde{\boldsymbol{u}}  -  \frac{\textbf{p}_i}{m_i}  \right) \cdot  d\textbf{S}_i \right].
\label{RTTmass} 
\end{align}
In a Lagrangian reference frame, the translational velocity of
CV surface must be equal to the molecular streaming velocity, {\it i.e.}, $ \tilde{\boldsymbol{u}}(\textbf{r}^\pm) = \boldsymbol{u}(\textbf{r}_i)$, so that,
\begin{align}
\displaystyle\sum_{i = 1}^{N} \left[ m_i \left( \boldsymbol{u}  -  \frac{\textbf{p}_i}{m_i}  \right) 
\cdot  d\textbf{S}_i \right] = -\displaystyle\sum_{i = 1}^{N} \overline{\textbf{p}}_i \cdot  d\textbf{S}_i. \nonumber 
\end{align}
The evolution of the peculiar momentum, $\boldsymbol{\chi} = \overline{\textbf{p}}_i$, in a moving reference frame is,
\begin{align}
 \frac{d}{d t}\displaystyle\sum_{i=1}^{N}  \overline{\textbf{p}}_i \vartheta_i
 =  \displaystyle\sum_{i = 1}^{N} \left[   \textbf{F}_i   \vartheta_i + \overline{\textbf{p}}_i \left( \boldsymbol{u} - \frac{\textbf{p}_i}{m_i} \right) \cdot  d\textbf{S}_i \right]
\nonumber \\
 =  \displaystyle\sum_{i = 1}^{N} \left[   \textbf{F}_i   \vartheta_i - \frac{\overline{\textbf{p}}_i  \overline{\textbf{p}}_i}{m_i} \cdot  d\textbf{S}_i \right]. \nonumber
\end{align}
Here an inertial reference frame has been assumed so that $d \overline{\textbf{p}}_i/dt = d\textbf{p}_i/dt = \textbf{F}_i$. For a simple case (e.g.\ one dimensional flow) it is possible to utilize a Lagrangian description by ensuring, $\tilde{\boldsymbol{u}}(\textbf{r}^\pm) = \boldsymbol{u}(\textbf{r}_i)$, throughout the time evolution. In more complicated cases, this is not always possible and the Eulerian description is generally adopted.

Next, a microscopic analogue to the macroscopic divergence theorem is derived for the generalized function, $\boldsymbol{\chi}$, 
\begin{align}
\int_V \displaystyle\sum_{i=1}^{N} \frac{\partial}{\partial \textbf{r}} \cdot  \bigg[ \boldsymbol{\chi} (\textbf{r}_i,\textbf{p}_i,t) \delta(\textbf{r}_i-\textbf{r}) \bigg] dV
\nonumber \\
=\int_V \displaystyle\sum_{i=1}^{N}  \boldsymbol{\chi} (\textbf{r}_i,\textbf{p}_i,t) \cdot \frac{\partial}{\partial \textbf{r}} \delta(\textbf{r}_i-\textbf{r}) dV. \nonumber
\end{align}
The vector derivative of the Dirac $\delta$ followed by the integral over volume results in,
\begin{align}
\int_V \frac{\partial}{\partial \textbf{r}} \delta(x_i-x)\delta(y_i-y)\delta(z_i-z)dV \;\;\;\;\;\;\;\;\;
\nonumber \\
=\begin{pmatrix}
  	\left[ \delta(x_i-x) H(y_i-y)  H(z_i-z) \right]_V   \\
 	\left[ H(x_i-x) \delta(y_i-y)  H(z_i-z) \right]_V  \\
 	\left[ H(x_i-x) H(y_i-y)  \delta(z_i-z) \right]_V
\end{pmatrix} \;\;\;\;\;\;\;\;\;\;
\nonumber \\
= \begin{pmatrix}
     \left[\delta(x_i-x^+)-\delta(x_i-x^-)\right]S_{xi}    \\
     \left[\delta(y_i-y^+)-\delta(y_i-y^-)\right]S_{yi}    \\
     \left[\delta(z_i-z^+)-\delta(z_i-z^-)\right]S_{zi}
 \end{pmatrix}
= d\textbf{S}_{i}, \nonumber
\end{align} 
where the limits of the cuboidal volume are, 
$\textbf{r}^+ = \textbf{r}+\frac{\Delta \textbf{r}}{2}$ and $\textbf{r}^- =\textbf{r}-\frac{\Delta \textbf{r}}{2}$. 
The mesoscopic equivalent of the continuum divergence theorem (Eq.\ (\ref{divergence})) is therefore,
\begin{align}
\int_V \frac{\partial}{\partial \textbf{r}} \cdot \displaystyle\sum_{i = 1 }^{N} \boldsymbol{\chi}   \delta(\textbf{r}_i - \textbf{r} )  dV   =\displaystyle\sum_{i = 1 }^{N} \boldsymbol{\chi} \cdot d\textbf{S}_{i}. \nonumber
\end{align}

\abovesectionspace
\section{Relation between Control Volume and Description at a Point}
\belowsectionspace
\label{sec:limitV0IK}
This  Appendix proves that the \citet{Irving_Kirkwood} expression for the flux at a point is 
the zero volume limit of the CV formulation. As in the continuum, the control volume equations 
at a point are obtained using the gradient operator in Eq.\ (\ref{definition_of_grad}).
the flux at a point can be shown by taking the zero volume limit of the gradient operator of 
Eq.\ (\ref{definition_of_grad}). 
Assuming the three side lengths of the control volume, $\Delta x, \Delta y$ and $\Delta z$, tend to zero 
and hence the volume, $\Delta V$, tends to zero,
 \begin{align}
\boldsymbol{\nabla} \cdot \rho \boldsymbol{u} = \lim_{\Delta x \rightarrow 0} \lim_{\Delta y \rightarrow 0} \lim_{\Delta z \rightarrow 0} \frac{1}{\Delta x \Delta y \Delta z} 
\nonumber \\
\times \displaystyle\sum_{i = 1 }^{N} \bigg\langle p_{ix} \frac{\partial \vartheta_i}{\partial x}
+ p_{iy}\frac{\partial \vartheta_i}{\partial y}
+ p_{iz} \frac{\partial \vartheta_i}{\partial z} ; \textit{f}  \bigg\rangle.
\label{limmassflux} 
 \end{align}
from Eq.\ (\ref{C2M_mass}).
For illustration, consider the $x$ component above, where
\begin{align}
\frac{\partial \vartheta_i}{\partial x} = \overbrace{\left[ \delta(x^+ - x_i)-\delta(x^- - x_i) \right]}^{x_{face}} S_{xi}.
\label{xfaceSi}
 \end{align}
Using the definition of the Dirac $\delta$ function as the limit of two slightly displaced Heaviside functions,
\begin{align}
 \delta(\xi)=\displaystyle\lim_{\Delta \xi\to0} \frac{H \left(\xi+\frac{\Delta \xi}{2}\right)-H\left(\xi-\frac{\Delta \xi}{2}\right)}{\Delta \xi},
\nonumber
\end{align}
the limit of the $S_{xi}$ term is,
\begin{align}
\lim_{\Delta y \rightarrow 0} \lim_{\Delta z \rightarrow 0} S_{xi} = \delta(y_i - y)\delta(z_i-z) \nonumber
 \end{align}
The $\Delta x \rightarrow 0$ limit for $x_{face}$ (defined in \eq{xfaceSi}) can be evaluated using L'H\^{o}pital's rule, combined with the property of the $\delta$ function,
\begin{align}
\frac{\partial}{\partial (\Delta \xi)} \delta\left(\xi-\frac{\Delta \xi}{2}\right) = - \frac{1}{2} \frac{\partial}{\partial \xi} \delta\left(\xi-\frac{\Delta \xi}{2}\right), \nonumber 
\end{align}
so that,
 \begin{align}
 \displaystyle\lim_{\Delta x\to0} x_{face} 
= \! \frac{\partial}{\partial x}  \delta\left(x - x_i\right).
 \nonumber
 \end{align}
Therefore, the limit of $\partial \vartheta_i/\partial x$ as the volume approaches zero is,
\begin{align}
\lim_{\Delta x \rightarrow 0} \lim_{\Delta y \rightarrow 0} \lim_{\Delta z \rightarrow 0} \frac{\partial \vartheta_i}{\partial x}
= \frac{\partial}{\partial x}  \delta\left( \textbf{r}_i - \textbf{r}\right),
\nonumber 
 \end{align}
Taking the limits for the $x$, $y$ and $z$ terms in \eq{limmassflux} yields the expected \citet{Irving_Kirkwood} definition of the divergence at a point,
 \begin{align}
\boldsymbol{\nabla} \cdot \rho \boldsymbol{u} = \displaystyle\sum_{i = 1 }^{N} \bigg\langle \frac{\partial}{\partial \textbf{r}} \cdot \textbf{p}_{i}  \delta(\boldsymbol{r_i} - \boldsymbol{r})  ; \textit{f}  \bigg\rangle. \nonumber
 \end{align}
This zero volume limit of the CV surface fluxes shows that the 
divergence of a Dirac $\delta$ function represents the flow of molecules 
over a point in space. The 
advection
and kinetic pressure at a point is, from \eq{KTeq},
\begin{align}
\boldsymbol{\nabla} \cdot \left[ \rho \boldsymbol{u} \boldsymbol{u} + \boldsymbol{\kappa} \right] = \displaystyle\sum_{i = 1 }^{N} \bigg\langle \frac{\partial}{\partial \textbf{r}} \cdot  \frac{\textbf{p}_i \textbf{p}_i }{m_i} \delta(\boldsymbol{r_i} - \boldsymbol{r})   ; \textit{f}  \bigg\rangle. \nonumber
\end{align} 
The same limit of zero volume for the surface tractions
defines 
the Cauchy stress. Using Eq.\ (\ref{definition_of_grad}) 
and taking the limit of Eq.\ (\ref{micro_face_stress}), written in terms of tractions,
 \begin{align}
\boldsymbol{\nabla} \cdot \boldsymbol{\sigma} \! =  \!\! \lim_{\Delta V \rightarrow 0} \! \frac{1}{\Delta V} \!\! \displaystyle\sum_{faces}^{6} \int_{S_{f}} \! \boldsymbol{\sigma} \cdot d\textbf{S}_{f} = \!\! \lim_{\Delta r_x \rightarrow 0} \lim_{\Delta r_y \rightarrow 0} \lim_{\Delta r_z \rightarrow 0} 
\nonumber \\
\times \left[  \frac{\textbf{T}_x^+ - \textbf{T}_x^-}{\Delta r_x} + \frac{\textbf{T}_y^+ - \textbf{T}_y^-}{\Delta r_y} + \frac{\textbf{T}_z^+ - \textbf{T}_z^-}{\Delta r_z} \right]. \nonumber
 \end{align}
For the $r_x^+$ surface, and taking the limits of  $\Delta r_y$ and $\Delta r_z$ using L'H\^{o}pital's rule,
 \begin{align}
\lim_{\Delta V \rightarrow 0} \frac{\textbf{T}_x^+}{\Delta r_x} 
= - \lim_{\Delta r_x \rightarrow 0} \frac{1}{2\Delta r_x} \displaystyle\sum_{i,j}^{N} \bigg\langle f_{\alpha ij} \varpi_{xyz}^+ ; \textit{f}  \bigg\rangle. \nonumber
 \end{align}
where $\varpi$ is 
 \begin{align}
\varpi_{\beta\kappa \gamma}^\dag \define  \left[H(r_\beta^\dag - r_{\beta j}) - H(r_\beta^\dag - r_{\beta i})\right] \; \; 
\nonumber \\
\times \delta\left(r_{\kappa} - r_{\kappa i} - \frac{r_{\kappa ij}}{r_{\beta ij}}\left( r_\beta^\dag - r_{\beta i} \right)\right) \; \;
\nonumber \\
\times \delta\left(r_{\gamma} - r_{\gamma i} - \frac{r_{\gamma ij}}{r_{\beta ij}}\left( r_\beta^\dag - r_{\beta i} \right)\right).
\label{pointonline}
 \end{align}
The indices $\beta, \kappa$ and $\gamma$ can be $x,y$ or $z$ and $\dag$ denotes the top surface ($+$ superscript), 
bottom surface ($-$ superscript) or CV center (no superscript). The $\varpi$ selecting function 
includes only the contribution to the stress when the line of interaction between $i$ and $j$ passes through 
the point $\textbf{r}^\dag$ in space. The difference between $\textbf{T}_x^+$ and $\textbf{T}_x^-$ tends to 
zero on taking 
the limit $\Delta r_x \to 0$,
so that L'H\^{o}pital's rule can be applied. Using the property,
\begin{align}
  \frac{\partial}{\partial (\Delta \xi)} \delta\left(\xi-\frac{1}{2}\Delta \xi\right) H\left(\xi-\frac{1}{2}\Delta \xi\right) 
\nonumber \\
= - \frac{1}{2} \frac{\partial}{\partial \xi} \delta\left(\xi-\frac{1}{2}\Delta \xi\right) H\left(\xi-\frac{1}{2}\Delta \xi\right), \nonumber  
\end{align}
then,
\begin{align}
\lim_{\Delta V \rightarrow 0} \frac{ \textbf{T}_x^+ - \textbf{T}_x^- }{\Delta r_x} 
= - \frac{1}{2} \displaystyle\sum_{i,j}^{N} \bigg\langle f_{\alpha ij} \frac{\partial \varpi_{xyz}}{\partial r_x}  ; \textit{f}  \bigg\rangle \!. \nonumber
\end{align}
where 
$r^+ \rightarrow r$ and $r^- \rightarrow r$. The $\varpi_{\beta \kappa \gamma}$ function 
is the integral between two molecules introduced in \eq{intritojds},
\begin{align}
\int\limits_{0}^{1}  \delta(\textbf{r} - \textbf{r}_i + s \textbf{r}_{ij}) ds = sgn \left(\frac{1}{r_{x ij}}\right) \frac{1}{|r_{x ij}|} 
\nonumber \\
\times \left[H(r_x - r_{x j}) - H(r_x - r_{x i})\right] \; \; 
\nonumber \\
\times \delta\left(r_{y} - r_{y i} - \frac{r_{y ij}}{r_{x ij}}\left( r_x - r_{x i} \right)\right) \; \;
\nonumber \\
\times \delta\left(r_{z} - r_{z i} - \frac{r_{z ij}}{r_{x ij}}\left( r_x - r_{x i} \right)\right). \nonumber
 \end{align}
where the sifting property of the Dirac $\delta$ function in the $r_x$ direction has been used to 
express the integral between two molecules in terms of the $\varpi_{x y z}$ function. Hence,
\begin{align}
 \int\limits_{0}^{1}  \delta(\textbf{r} - \textbf{r}_i + s \textbf{r}_{ij}) ds = \frac{\varpi_{x y z}}{r_{x ij}}. \nonumber
 \end{align}
As the choice of shifting direction is arbitrary, use of $r_y$ or $r_z$ in the above treatment would result in 
$\varpi_{y z x}$ and $\varpi_{z x y}$, respectively. Therefore, \eq{divofstress}, 
without the volume integral, can be expressed as,
 \begin{align}
\frac{1}{2}\displaystyle\sum_{i,j}^{N} \bigg\langle f_{\alpha ij} r_{\beta ij} \frac{\partial}{\partial r_{\beta}}  \int\limits_{0}^{1}  \delta(\textbf{r} - \textbf{r}_i + s \textbf{r}_{ij}) ds ; \textit{f}  \bigg\rangle 
\nonumber \\ 
= \frac{1}{2}   \displaystyle\sum_{i,j}^{N} \bigg\langle   f_{ij\alpha} \bigg[ \frac{\partial \varpi_{xyz}}{\partial r_x} \; + \, \frac{\partial \varpi_{yxz}}{\partial r_y} \,  +  \frac{\partial \varpi_{zxy}}{\partial r_z}  \bigg] ; \textit{f}  \bigg\rangle. \nonumber
 \end{align}
As \eq{divofstress} is equivalent to the \citet{Irving_Kirkwood} stress of \eq{stress_equality}, 
the Irving Kirkwood stress is recovered in the limit that the CV tends to zero volume. \\
This Appendix has proved therefore that in the limit of zero control volume, the molecular CV Eqs.\ (\ref{CVmass_eqn}) 
and (\ref{mesoevo_traction}) recover the description at a point in the same limit that the continuum 
CV Eqs.\ (\ref{BofmEqn2}) and (\ref{BofMEqn2}) tend to the differential continuum equations. 
This demonstrates that the molecular CV equations presented here are the molecular scale equivalent of the continuum CV equations. 

\abovesectionspace
\section{Relationship between Volume Average and Method Of Planes Stress}
\belowsectionspace
\label{sec:lineplane}

This Appendix gives further details of the derivation of the Method Of Planes 
form of stress from the Volume Average form. 
Starting from Eq.\ (\ref{divofstress}) written in terms of the CV function for an integrated volume,
 \begin{align}
-\displaystyle\sum_{faces}^{6} \int_{S_{f}} \boldsymbol{\sigma} \cdot d\textbf{S}_{f} =  \frac{1}{2}\displaystyle\sum_{i,j}^{N} \bigg\langle   \textbf{f}_{ij} \textbf{r}_{ij} \cdot  \int\limits_{0}^{1} \frac{\partial  \vartheta_s }{\partial \textbf{r}}  ds ; \textit{f}  \bigg\rangle \; 
\nonumber \\
 = \frac{1}{2}\displaystyle\sum_{i,j}^{N} \bigg\langle   \textbf{f}_{ij} \int\limits_{0}^{1}  \left[ x_{ij}  \frac{\partial  \vartheta_s }{\partial x}  + y_{ij} \frac{\partial  \vartheta_s }{\partial y}  + z_{ij}   \frac{\partial  \vartheta_s }{\partial z}  \right] ds ; \textit{f}  \bigg\rangle.
\label{VA_start}
 \end{align}
Taking only the $x$ derivative above,
 \begin{align}
x_{ij} \frac{\partial \vartheta_s }{\partial x} 
=   x_{ij} \big[  \overbrace{\delta(x^+ - x_i + s x_{ij})}^{x^+_{face}} \;\;\;\;\;\;\;\;
\nonumber \\
- \delta(x^- - x_i + s x_{ij}) \big] G(s)
\label{xfacecrossing}
 \end{align}
where $G(s)$ is,
 \begin{align}
G(s) \define \left[ H(y^+ - y_i + s y_{ij}) - H(y^- - y_i + s y_{ij}) \right] \; \,
\nonumber \\
\times\left[ H(z^+ - z_i + s z_{ij}) - H(z^- - z_i + s z_{ij}) \right]. \nonumber
 \end{align}
As $\delta(ax) = \frac{1}{|a|}\delta(x)$ the $x_{ij} x^+_{face} G(s)$ term in \eq{xfacecrossing} can be expressed as,
 \begin{align}
 x_{ij} x^+_{face} G(s) = \frac{ x_{ij}}{|x_{ij}|} \delta \left(\frac{x^+ - x_i}{x_{ij}} + s \right) G(s).
 \end{align}
The integral can be evaluated using the sifting property of the Dirac $\delta$ function \citep{Thankoppan} as follows,
 \begin{align}
 \int\limits_{0}^{1} x_{ij}   x^+_{face} G(s) ds = \frac{x_{ij}}{|x_{ij}|} \int\limits_{0}^{1}  \! \delta \! \left(\frac{x^+ - x_i}{x_{ij}} + s \right)G(s)ds
 \nonumber \\
= sgn(x_{ij}) \bigg[  H\left(\frac{x^+ - x_j}{x_{ij}}\right) - H\left(\frac{x^+ - x_i}{x_{ij}}\right) \bigg]S_{xij}^+. \nonumber
 \end{align}
where the signum function, $sgn(x_{ij})\define x_{ij}/|x_{ij}|$. 
The  $S_{xij}^+$ term is the value of $s$ on the cube surface, \\ $S_{xij}^+ = G\left(s=-\frac{x^+ - x_i}{x_{ij}}\right)$ which is,
 \begin{align}
S_{xij}^+ \define \left[H\left(y^+ - y_i - \frac{y_{ij}}{x_{ij}}\left( x^+ - x_i \right)\right)  \; \; \;
\nonumber \nl
- H\left(y^- - y_i - \frac{y_{ij}}{x_{ij}}\left( x^+ - x_i \right)\right)\right] \; \,
\nonumber \\
\times \left[H\left(z^+ - z_i - \frac{z_{ij}}{x_{ij}}\left( x^+ - x_i \right)\right) \; \; \;
\nonumber \nl
- H\left(z^- - z_i - \frac{z_{ij}}{x_{ij}}\left( x^+ - x_i \right)\right)\right].
\label{Sijdef}
 \end{align}

The definition $S_{xij}^+$ (analogous to $S_{xi}$ in Eq.\ (\ref{Six})) has been introduced 
as it 
filters out those $ij$ terms where
the point of intersection of line $r_{ij}$ and plane $x^+$ has $y$ and $z$ 
components between the limits of the cube surfaces. 
The corresponding terms, $S_{ij\alpha}^\pm$, are defined for  $\alpha = \{y,z\}$. 
Taking $H(0) = \frac{1}{2}$, the Heaviside function can be rewritten as $H(ax) = \frac{1}{2}\left(sgn(a)sgn(x)-1\right)$, and,
 \begin{align}
H\left(\frac{x^+ - x_j}{x_{ij}}\right) - H\left(\frac{x^+ - x_i}{x_{ij}}\right)
\nonumber \\
\!\!\!=\frac{1}{2}sgn\left(\!\frac{1}{x_{ij}}\!\right)\left[sgn(x^+ - x_j) - sgn(x^+ - x_i)\right], \nonumber
 \end{align}
so the expression, $x_{ij} x^+_{face} G(s)$ in \eq{xfacecrossing} becomes,
 \begin{align}
 x_{ij} \int\limits_{0}^{1}  x^+_{face} G(s) ds  = \frac{1}{2}  sgn(x_{ij}) sgn\left(\frac{1}{x_{ij}}\right)
\nonumber \\  
\!  \times \left[sgn(x^+ - x_j) - \! sgn(x^+ - x_i)\right] \!  \! S_{xij}^+. \nonumber
 \end{align}
The signum function, $sgn\left(\frac{1}{x_{ij}}\right)$, cancels the one obtained from integration along $s$, $sgn(x_{ij})$.
The expression for the $x^+$ face is therefore,

 \begin{align}
-\int_{S^+_{x}} \boldsymbol{\sigma} \cdot d\textbf{S}_{S^+_{x}} = \frac{1}{2}\displaystyle\sum_{i,j}^{N} \bigg\langle   \textbf{f}_{ij}  x_{ij} \int\limits_{0}^{1}  x_{face}^+ G(s) ds ; \textit{f}  \bigg\rangle
\nonumber \\
 =\frac{1}{4} \displaystyle\sum_{i,j}^{N} \bigg\langle  \textbf{f}_{ij} \left[sgn(x^+ - x_j) - sgn(x^+ - x_i)\right]S_{xij}^+   ; \textit{f}  \bigg\rangle \nonumber
 \end{align}
Repeating the same process for the other faces allows \eq{VA_start} to be expressed as,

 \begin{align}
 \displaystyle\sum_{faces}^{6} \int_{S_{f}} \boldsymbol{\sigma} \cdot d\textbf{S}_{f} = -  \frac{1}{2}\displaystyle\sum_{i,j}^{N} \bigg\langle   \textbf{f}_{ij} \textbf{r}_{ij} \cdot  \int\limits_{0}^{1} \frac{\partial  \vartheta_s }{\partial \textbf{r}}  ds ; \textit{f}  \bigg\rangle
\nonumber \\
= -\frac{1}{4}\displaystyle\sum_{i,j}^{N} \bigg\langle   \textbf{f}_{ij} \displaystyle\sum_{\alpha=1}^{3} \tilde{n}_{\alpha} \left[ dS_{\alpha ij}^+ - dS_{\alpha ij}^- \right] ; \textit{f}  \bigg\rangle
,\nonumber
 \end{align}
where  $dS_{\alpha ij}^\pm \define \frac{1}{2}\left[sgn(r^\pm_\alpha - r_{\alpha j}) - sgn(r^\pm_\alpha - r_{\alpha i})\right] S_{\alpha ij}^\pm$ and $\tilde{n}_\alpha \define sgn(r_{\alpha ij}) sgn\left(\frac{1}{r_{\alpha ij}}\right) = [1 \; 1 \; 1]$. This is the force over the CV surfaces, \!\!\!\! \eq{micro_face_stress}, in section \ref{sec:pressure}.

To verify the interpretation of $S_{xij}^+$ used in this work, consider the vector equation for the point of intersection of a line and a plane in space. The equation for a vector $\textbf{a}$ between $\textbf{r}_i$ and $\textbf{r}_j$ is defined as $\textbf{a} = \textbf{r}_i - s \frac{\textbf{r}_{ij}}{|\textbf{r}_{ij}|}$. The plane containing the positive face of a cube is defined by $ \left( \textbf{r}^+ - \textbf{p} \right) \cdot \textbf{n}$ where $\textbf{p}$ is any point on the plane and $\textbf{n}$ is normal to that plane. By setting $\textbf{a}=\textbf{p}$ and upon rearrangement of $\left( \textbf{r}^+ - \textbf{r}_i + s \frac{\textbf{r}_{ij}}{|\textbf{r}_{ij}|} \right) \cdot \textbf{n}$, the value of $s$ at the point of intersection with the plane is,
\begin{align}
s= -\frac{\left( \textbf{r}^+ - \textbf{r}_i \right) \cdot \textbf{n} }{\frac{\textbf{r}_{ij}}{|\textbf{r}_{ij}|} \cdot \textbf{n}}, \nonumber
\end{align}
The point on line $\textbf{a}$ located on the plane is,
\begin{align}
\textbf{a}_{p}^+ \define \textbf{r}_i + \textbf{r}_{ij} \left[ \frac{\left( \textbf{r}^+ - \textbf{r}_i \right) \cdot \textbf{n} }{\textbf{r}_{ij} \cdot \textbf{n}} \right]. \nonumber
 \end{align}
Taking $\textbf{n}$ as the normal to the $x$ surface, i.e.\\  $ \textbf{n} \rightarrow \textbf{n}_x = [1,0,0]$, then,
 \begin{align}
x_{\alpha p}^+ =  \begin{pmatrix}
   x_{x p}^+ \\
   x_{y p}^+  \\
   x_{z p}^+ 
 \end{pmatrix} =
 \begin{pmatrix}
   x^+ \\
  y_i +  \frac{y_{ij}}{x_{ij}} \left( x^+ - x_i \right)   \\
  z_i +  \frac{z_{ij}}{x_{ij}} \left( x^+ - x_i \right)
 \end{pmatrix} \nonumber 
\end{align}
written using index notation with $\alpha = \{ x,y,z \}$. The vector $\textbf{x}_{p}^+$ is the point of intersection of line $\textbf{a}$ with the $x^+$ plane. 
A function to check if the point $\textbf{x}_{p}^+$ on the plane is located on the region between $y^\pm$ and $z^\pm$, would use Heaviside functions and is similar to the form of \eq{Six},
\begin{align}
S_{xij}^+ = \left[ H\left(y^+ - \textbf{x}_{yp}^+ \right) -H\left(y^- - \textbf{x}_{yp}^+ \right) \right] \;\;
\nonumber \\
\times \left[ H\left(z^+ - \textbf{x}_{zp}^+\right)-H\left(z^- -\textbf{x}_{zp}^+\right) \right], \nonumber
\end{align}
which is the form obtained in the text by direct integration of the expression for stress, \textit{i.e.} Eq.\ (\ref{Sijdef}).


%

\end{document}